\documentclass[sigconf, nonacm]{acmart}

\newcommand\vldbdoi{XX.XX/XXX.XX}

\newcommand\vldbavailabilityurl{URL_TO_YOUR_ARTIFACTS}
\newcommand\vldbpagestyle{plain}
\PassOptionsToPackage{dvipsnames}{xcolor}
\usepackage[dvipsnames]{xcolor}
\usepackage{hyperref}
\usepackage{listings}
\usepackage{listingsutf8}
\usepackage{makecell}
\usepackage{color}
\usepackage[T1]{fontenc}
\usepackage[utf8]{inputenc}
\usepackage{subcaption}
\usepackage{graphicx}
\usepackage[ngerman, english]{babel}
\usepackage{bibgerm}
\usepackage{balance}
\usepackage{enumitem}
\usepackage{csquotes}
\usepackage{xcolor}
\DeclareCaptionType[fileext=lol]{listing}
\usepackage{verbatim}
\usepackage{textcomp}
\usepackage{tikz}
\usepackage{multirow}
\usepackage{tabularx}
\usepackage[ruled,vlined]{algorithm2e}
\usepackage{amsmath}
\usepgflibrary{plotmarks}
\usepgflibrary[plotmarks]
\usetikzlibrary{plotmarks}
\usetikzlibrary[plotmarks]
\usepackage{fvextra}
\usepackage{pgfplots}
\usepgfplotslibrary{groupplots}
\pgfplotsset{compat=newest}
\usepackage{pgfplotstable}

\definecolor{codebackground}{gray}{0.90}
\lstdefinestyle{code}{
    frame=tb,
    framerule=0pt,
    backgroundcolor=\color{gray!20},
    commentstyle=\color{Green},
    keywordstyle=\color{magenta},
    numberstyle=\tiny\color{gray},
    stringstyle=\color{purple},
    basicstyle=\ttfamily\footnotesize,
    breakatwhitespace=false,
    breaklines=true,
    captionpos=b,
    keepspaces=true,
    numbers=left,
    numbersep=5pt,
    showspaces=false,
    showstringspaces=false,
    showtabs=false,
    tabsize=2
}

\newcommand{\dsname}{\textsf{full-named}}
\newcommand{\dsnoname}{\textsf{full-unnamed}}
\newcommand{\deslice}{\textsf{de-slice}}
\newcommand{\usslice}{\textsf{us-slice}}

\newcommand{\ruslice}{\textsf{ru-slice}}

\newcommand{\library}{\textsf{library}}
\newcommand{\shoppingcenter}{\textsf{shopping-center}}

\begin{document}

\title{Clean Me If You Can: A Large Collection of Real-World Addresses for Data Cleaning Benchmarking [Experiment, Analysis \& Benchmark]}

\renewcommand{\shorttitle}{Clean Me If You Can: A Large Collection of Real-World Addresses for Data Cleaning Benchmarking}

\author{Fatemeh Ahmadi$^1$, Tobias Bernhard$^1$, Mohamed Abdelmaksoud$^1$, Luca Zecchini$^1$,\\ Tilmann Rabl$^2$, Ziawasch Abedjan$^1$}
\affiliation{%
  \institution{$^1$BIFOLD \& TU Berlin, $^2$Hasso Plattner Institute, Potsdam}
}
\email{{f.ahmadi, mohamed, luca.zecchini, abedjan}@tu-berlin.de, tilmann.rabl@hpi.de, to-tb@gmx.de}

\renewcommand{\shortauthors}{Ahmadi et al.}

\begin{abstract}
There has been extensive research on automating and scaling data cleaning, i.e., the detection and correction of erroneous values in tabular data. Yet, existing approaches often perform well only within controlled environments. One of the major bottlenecks in data cleaning research is the lack of real-world datasets. In this paper, we address this gap by providing a large, dirty dataset with postal entries and their corresponding ground truth. We discuss the design decisions and challenges for obtaining the dataset. We demonstrate the limitations of existing cleaning approaches when faced with our proposed datasets and derive guidelines for future research.
\end{abstract}

\maketitle
\textit{Preprint. Submitted for review.}
\vspace{0.5em}

\pagestyle{\vldbpagestyle}
\begingroup\small\noindent\raggedright
\endgroup

\ifdefempty{\vldbavailabilityurl}{}{
\vspace{.3cm}
\begingroup\small\noindent\raggedright\textbf{Artifact Availability:}\\
The source code, data, and other artifacts have been made available at \url{https://github.com/D2IP-TUB/Clean-Me-If-You-Can}}.
\endgroup

\setcounter{section}{0}
\section{Introduction}
\label{sec:introduction}

Ensuring high-quality data is a cumbersome process that requires numerous exhaustive human cycles~\cite{DBLP:journals/sigmod/MohammedEHNS25}. Data cleaning research aims to automate this task, proposing a large number of rule-based~\cite{DBLP:journals/pvldb/RezigOAEMS21}, learning-based~\cite{DBLP:journals/pvldb/MahdaviA20}, and, more recently, LLM-driven~\cite{DBLP:conf/icde/WuYZMNXZY25} approaches.
Although individual papers often report improved results, industrial adoption remains a challenge. This is partially due to the fact that many approaches are not scalable, or they target specific error types and distributions that do not reflect the complexity of the real world~\cite{DBLP:journals/jdiq/JungJCB25}.

One impeding factor to data cleaning research is the absence of real-world benchmark datasets to guide the research progress. For evaluating a cleaning algorithm, two versions of a dataset are necessary: the so-called \emph{dirty} version, which requires cleaning, and the \emph{ground truth}, which contains all records in the desired cleaned format.
Generally, there are three possible approaches to obtain such benchmark datasets: \emph{(i)}~a supposedly clean dataset is dirtied using error generators~\cite{DBLP:journals/pvldb/ArocenaGMMPS15, DBLP:journals/jdiq/JungJCB25}, \emph{(ii)}~a dirty dataset is manually curated to obtain the corresponding ground truth~\cite{AbedjanCDFIOPST16}, or \emph{(iii)}~the ground truth does not exist and the success of a downstream task, such as inference accuracy, is used as a proxy for cleanliness~\cite{DBLP:journals/is/MohammedBFINNPNH25,DBLP:journals/pacmmod/SiddiqiKB23,DBLP:conf/icde/LiRBZCZ21}.

Each of these approaches is generally viable but has specific limitations. Error generators are useful for benchmarking algorithm behavior in controlled environments, but they often rely on randomness or explicit patterns, in contrast to latent patterns that manifest in the real world.
Tabular errors result from systematic flaws in data collection, such as human typos, flawed extractors, outdated values, etc.
In contrast, when building a benchmark from naturally dirty data, the central challenge is constructing ground truth at scale. This is either feasible for small-scale datasets or for very obvious errors that can be cleaned with very simple heuristics. Finally, using downstream task performance circumvents the need for obtaining ground truth.
However, downstream proxy scores are not always guiding towards a clean version of a dataset, as proxy scores such as downstream inference accuracy might even improve through artificial noise~\cite{DBLP:journals/is/MohammedBFINNPNH25}.

To address the lack of realistic large-scale benchmark datasets, we put our efforts into providing researchers and practitioners with a real-world dataset that can be used for benchmarking data cleaning solutions.
Cleaning postal information has been a primary use case in data quality assessment, as companies try to retain curated information about (prospective) customers~\cite{digital4040051}, suppliers, and partners~\cite{Leeuw1992DataQI}. The famous case of the 6.8 billion pieces of mail that could not be delivered as addressed~\cite{usps}
further underpins the importance of data quality in such a context.
Despite this relevance, a real-world benchmark address dataset has so far been missing. Prior research was conducted on either proprietary address datasets or a rather limited restaurant dataset~\cite{AbedjanCDFIOPST16}.

In this paper, we present a dataset that contains real addresses of organizations across the world with real-world data quality problems.
To obtain this dataset, we exploited the Web Data Commons (WDC) corpus~\cite{DBLP:conf/www/BrinkmannPB23} and used Web services to collect postal information on public entities.
The resulting dataset contains up to 9,317,886 records, which exhibit data quality issues that stem from language inconsistencies, inaccuracies, obsolescence, and extraction limitations~\cite{DBLP:journals/pvldb/AbedjanAOPS15}. We also provide further variations of the dataset that differ in terms of geographical scope, type of entities, and cleanliness.

We show that a rather low-dimensional real-world dataset is already challenging for state-of-the-art research prototypes, opening avenues for further research in error detection and correction. In summary, we make the following contributions.

\begin{enumerate}[leftmargin=*]
    \item We present a real-world benchmark dataset for scalable data cleaning that contains real errors. We also provide the corresponding ground truth derived from geocoding services. We share our process for obtaining this benchmark dataset. The process is fully reproducible and our dataset and extraction code are available online.

    \item We conduct an extensive evaluation that highlights the limitations of existing approaches for both error detection and correction. None of the evaluated data cleaning approaches can reliably clean the dataset or even its slices. Learning-based approaches generally achieve strong effectiveness, but they also expose clear scalability limitations.

    \item We identify and discuss future directions for data cleaning research, supported by empirical evidence from our benchmark and evaluation. In particular, our results show the need for approaches that can benefit from corrections beyond the input dataset, balance classic rule-based, learning-based, and LLM-driven strategies, and scale not only with respect to the number of rows and columns, but also across different error rates and error distributions.

\end{enumerate}

\section{Related Work}

Benchmark datasets for data cleaning appear in three adjacent lines of research.

\subsubsection*{Data Cleaning Research}
Most real-world benchmark datasets have been established on-the-fly while proposing novel data cleaning techniques.
This is the case for datasets like Hospital~\cite{DBLP:conf/icde/ChuIP13}, Flights~\cite{DBLP:journals/pvldb/LiDLMS12}, Rayyan~\cite{ouzzani2016rayyan}, or Beers~\cite{beers_dataset}.
In 2015, Abedjan et al.~\cite{AbedjanCDFIOPST16} conducted an evaluation of state-of-the-art error detection techniques from that time.
To focus on real-world errors, the authors mostly relied on proprietary datasets.
Ni et al.\ extended the evaluation to data repair~\cite{DBLP:journals/pvldb/NiMZWLY24}.
Overall, most of the existing datasets are either small or contain errors that are easy to identify and fix. For example, Flights contains duplicate records acquired from different sources and
presents a strong correlation between the source of a record and its quality. Thus, errors can be easily detected and corrected using records from higher-quality sources.

\subsubsection*{Error Generators}
Several benchmarking papers provided error generation tools for datasets.
BART~\cite{DBLP:journals/pvldb/ArocenaGMMPS15} represents one of the earliest efforts to systematically generate data errors for benchmarking data cleaning systems, enabling controlled injection of diverse error types into clean databases. It remains widely used for constructing synthetic benchmarks~\cite{DBLP:conf/edbt/0003HS23}. JENGA~\cite{DBLP:conf/edbt/SchelterRB21} is a framework that includes a data corruption module to study the impact of data quality issues on downstream machine learning tasks.
GouDa~\cite{DBLP:conf/deem/RestatBCS22} is a data generation tool that creates datasets with specific error types without requiring an initial clean table.
More recently, Jung et al.~\cite{DBLP:journals/jdiq/JungJCB25} argued that errors in real-world data often depend on the data itself, whereas many existing benchmarks inject errors independently of the underlying data. They proposed a latent-factor model that captures dependencies between data values and error occurrences through shared latent variables.
While these frameworks facilitate controlled evaluations, they still produce synthetic errors with a set of predefined types rather than naturally occurring ones, limiting their ability to reflect the full diversity of real-world data issues.

\subsubsection*{Data Cleaning for Machine Learning}
Data cleaning can be an operator within a data science pipeline~\cite{DBLP:conf/edbt/SchelterRB21}. Thus, one can measure the effectiveness of cleaning using downstream model performance as a proxy.
Karlaš et al.\ proposed CPClean~\cite{DBLP:journals/pvldb/KarlasLWGC0020}, a data cleaning framework built on top of machine learning pipelines, and evaluated the impact of cleaning in the context of \textit{k}-nearest neighbor classifiers.
Abdelaal et al.\ proposed REIN~\cite{DBLP:conf/edbt/0003HS23}, a benchmark framework that enables data cleaning evaluation for downstream machine learning tasks, such as regression, classification, and clustering. While using downstream accuracy is viable in some scenarios, it does not serve for general-purpose data cleaning assessment.
Indeed, it can be misleading, as the downstream performance optimum often differs from the performance on the actual ground truth data~\cite{DBLP:journals/is/MohammedBFINNPNH25}.

\emph{With our benchmark dataset}, we expand existing test cases in two directions: we provide a \emph{very large real-world dataset with ground truth} and we curate smaller subsets for specific analysis tasks. Our dataset is significantly larger than existing academic datasets, showcasing the limitations of current cleaning methods.

\section{Address Dataset}

In this section, we iterate through the requirements we expect a benchmark dataset for data cleaning to meet and introduce the design of the targeted schema.

\subsection{Dataset Requirements}
\label{sec:dataset:requirements}
The main goal of this benchmark dataset is to support data cleaning research. We therefore identify several requirements that the final benchmark should fulfill.

    \textit{(1) Relevant domain.}
    The dataset should belong to a common domain for data cleaning and the errors should closely mirror those present in real-world data.

    \textit{(2) Content variations.}
    The dataset should capture a diverse set of data relationships at row and column levels.
    Thus, it should exhibit redundancy and conflicting records, different data types, as well as both related and independent attributes.

    \textit{(3) Error diversity.}
    The data should encompass various error types, such as missing values, inconsistent formatting, contradictory data, misspellings, or nonsensical values.
    Some errors should be easy to fix without additional context, while others might require consideration of attribute relationships and external knowledge.

    \textit{(4) Ground truth labels.}
    We are interested in how well a system can detect and correct errors.
    With the ground truth data for each record it is possible to determine which values are erroneous and what the correct value should be, enabling the calculation of metrics such as precision and recall.

    \textit{(5) Provenance.}
    The dataset should track the origin of each record. This enables transparent inspection of benchmark construction and allows future users to verify, reproduce, or extend the dataset.

\subsection{Domain: Postal Addresses}
\label{sec:requirement:design}

For our benchmark dataset, we choose the domain of postal information.
Postal address data is prevalent in a wide range of real-world applications, such as e-commerce or logistics.
It is one of the most common types of data that is subject to cleaning~\cite{AbedjanCDFIOPST16}.
While prevalent, address information is often challenging, as formats do not follow a single, universally adopted standard.

As a result, the address structure can vary considerably depending on local postal conventions and cultural differences.
For instance, the house number typically precedes the street name in US addresses (e.g., \textit{17 Main Street}) and follows it in German addresses (e.g., \textit{Hauptstraße 17}).
Addresses commonly include abbreviations, and even punctuation is often used inconsistently, separating address components through commas, whitespaces, or dashes.
Further, address styles can be heavily location-specific.
For example, the old town of Mannheim in Germany employs a distinctive addressing scheme using block-level identifiers and house numbers, while other districts of Mannheim adhere to the conventional German format based on street name and house number.
Finally, postal data often exhibit missing values: while fields such as street or locality are integral to an address, other fields such as region or country are frequently missing.

\section{Dataset Creation}
\label{sec:dataset}

Choosing the postal domain for benchmarking also has practical advantages, as the data and ground truth are generally accessible and understandable by any data engineer.
Yet, as we will describe in this section, the scalable extraction and annotation of addresses requires a careful process for searching and postprocessing.
A rich source for arbitrarily created datasets that are operational is the Web. Thus, one could extract visible addresses in regular Web dumps, such as the Web Data Commons (WDC) corpus~\cite{DBLP:conf/www/BrinkmannPB23}. At the same time, there are several commercial services that provide clean address records. The process can be divided in six major steps that are introduced below and detailed and justified in the following subsections.

\begin{enumerate}[leftmargin=*]

\item \textbf{Download WDC N-Quads.}
The WDC project extracts structured data from Common Crawl\footnote{\url{https://commoncrawl.org}} and publishes pre-extracted RDF statements in N-Quads format~\cite{DBLP:conf/www/BrinkmannPB23}. We downloaded all class-specific Schema.org subsets\footnote{\url{https://webdatacommons.org/structureddata/2024-12/stats/schema_org_subsets.html}} that contain \textsf{PostalAddress} annotations\footnote{\url{https://schema.org/PostalAddress}}, covering types of points of interest (POIs) such as \textit{Library}, \textit{Hotel}, and \textit{Hospital}, as well as the \textsf{html-mf-adr} subset from the Microformats source\footnote{\url{https://webdatacommons.org/structureddata/2024-12/stats/stats.html}}.\\
$\quad\rightarrow\;$ 12.9B N-Quads from 22 Schema.org subsets\\
$\quad\rightarrow\;$ 41M N-Quads from the \textsf{html-mf-adr} subset

\item \textbf{Extract address information.}
We parsed the downloaded N-Quads to identify address entities and reconstruct them as dataset records. We extracted \textsf{schema:PostalAddress} entities from the class-specific Schema.org subsets and \textsf{vcard:Address} entities from the \textsf{html-mf-adr} subset.
After collecting its fields, each extracted address entity is represented as one \emph{dirty record}, as it may contain errors.\\
$\quad\rightarrow\;$ 247.9M dirty records from \textsf{schema:PostalAddress} entities\\
$\quad\rightarrow\;$ 11.8M dirty records from \textsf{vcard:Address} entities

\item \textbf{Filter out unverifiable addresses.}
We discarded dirty records that provided neither street name, nor locality, nor postal code, as their information was insufficient for geocoding validation.\\
$\quad\rightarrow\;$ 219.9M dirty records from \textsf{schema:PostalAddress} entities\\
$\quad\rightarrow\;$ 8.5M dirty records from \textsf{vcard:Address} entities

\item \textbf{Retrieve reference addresses.}
We searched for corresponding reference addresses using a geocoding service. After matching and validation, we treat these reference addresses as \emph{ground-truth records}. The matching and validation procedure is discussed in detail in Section~\ref{sec:data_validation}.\\
$\quad\rightarrow$ 29.7M pairs of dirty and ground-truth records

\item \textbf{Align the schema for dirty and ground-truth records.}
We mapped the attributes of dirty and ground-truth records to a common schema and excluded pairs in which the ground-truth record contained imprecise or incomplete information.
We also removed exact duplicates, i.e., records that appeared multiple times in the exact same format on the same Web page.\\
$\quad\rightarrow\;$ \textbf{\dsnoname{} address dataset}: 9,317,886 dirty records with the corresponding ground truth (without names)

\item \textbf{Filter out entity mismatches.}
We ensured that the name referred to the same real-world entity in both the dirty record and the corresponding ground truth.\\
$\quad\rightarrow\;$ \textbf{\dsname{} address dataset}: 4,338,109 dirty records with the corresponding ground truth (with names).

\end{enumerate}

We further refined the \dsname{} dataset into several variations.
A detailed overview of all datasets and their key properties is provided in Table~\ref{tab:facility-error-summary}.
Next, we describe each step in the dataset generation process in further detail.

\subsection{Dirty Data Collection}
\label{sec:data_collection}

To obtain addresses in their dirty version, we used structured data pre-extracted from Common Crawl snapshots and publicly released as RDF N-Quads by the WDC project~\cite{DBLP:conf/www/BrinkmannPB23}.
We used the WDC release for December 2024, the most recent at the time of data collection.

\subsubsection{Address Extraction}
\label{sec:address_extraction}

We extracted records containing address information from two complementary WDC sources: Schema.org and Microformats.

\noindent\textbf{Class-specific Schema.org subsets.}
WDC extracts Schema.org annotations from Web pages and publishes them in
class-specific subsets according to the annotated entity type, such as
\textit{Library}, \textit{Hotel}, and \textit{Hospital}.
We downloaded all 22 subsets that contain entities with \textsf{schema:PostalAddress}
annotations, comprising 12.9B RDF statements in N-Quads format.
In this format, each line encodes a single
(\textsf{subject}, \textsf{predicate}, \textsf{object}, \textsf{graph IRI}) tuple,
where \textsf{graph IRI} represents the source page URL.
Each address is therefore distributed across multiple lines, one per field, as shown in Listing~\ref{lst:nquads_example}.
For each node typed as \textsf{schema:PostalAddress}, we extracted its address fields, including \textsf{streetAddress}, \textsf{postalCode}, \textsf{addressLocality}, \textsf{addressRegion}, and \textsf{addressCountry}.
We then searched for an entity whose \textsf{schema:address} property pointed to that address node. From this entity, we extracted its \textsf{schema:name} property. In Listing~\ref{lst:nquads_example}, \texttt{\_:b1} is the address node, and \texttt{\_:b0} is the \textsf{LocalBusiness} entity whose address is \texttt{\_:b1}.
Listing~\ref{lst:extracted_record} shows a simplified extracted address with selected fields.

\begin{listing}[t]
\begin{lstlisting}[style=code, numbers=none, gobble=0]
_:b0  rdf:type           schema:LocalBusiness    <example.de/page> .
_:b0  schema:name        "NAME"                  <example.de/page> .
_:b0  schema:address     _:b1                    <example.de/page> .
_:b1  rdf:type           schema:PostalAddress    <example.de/page> .
_:b1  schema:streetAddress     "STREET"          <example.de/page> .
_:b1  schema:addressLocality   "CITY"            <example.de/page> .
\end{lstlisting}
\vspace{-10pt}
\caption{N-Quads example from the \texttt{LocalBusiness} Schema.org subset. The address node
(\texttt{\_:b1}) is linked to its parent entity (\texttt{\_:b0}) via
\texttt{schema:address}.}
\label{lst:nquads_example}
\end{listing}

\begin{listing}[t]
\begin{lstlisting}[language=Python, style=code, numbers=none, gobble=0]
{
  "address": {
    "streetAddress": "STREET",
    "addressLocality": "CITY",
  },
  "name": "NAME",
  "graph_iri": "https://example.de/page"
}
\end{lstlisting}
\vspace{-10pt}
\caption{Address extracted from the example in Listing~\ref{lst:nquads_example}.}
\label{lst:extracted_record}
\end{listing}

\noindent\textbf{\textsf{html-mf-adr} Microformats subset.}
WDC also provides a format-specific dump for the \texttt{html-mf-adr} microformat,
where address fields are encoded as HTML class names and extracted
into the W3C vCard vocabulary, as shown in Listing~\ref{lst:vcard_nquads}.
This subset covers 41M RDF statements in N-Quads format across nine shards.
We remapped vCard predicates to Schema.org field names to ensure a uniform structure across both sources.

This extraction phase yielded 247.9M \textsf{schema:PostalAddress} entities from the Schema.org subsets and 11.8M \textsf{vCard:Address} entities from the \textsf{html-mf-adr} subset. After collecting its fields, we represent each extracted address entity as one dirty record.

\begin{listing}[t]
\begin{lstlisting}[style=code, numbers=none, gobble=0]
_:b0  rdf:type                vcard:Address     <example.com/page> .
_:b0  vcard:street-address    "STREET"          <example.com/page> .
_:b0  vcard:locality          "CITY"            <example.com/page> .
_:b0  vcard:country-name      "COUNTRY"         <example.com/page> .
\end{lstlisting}
\vspace{-10pt}
\caption{N-Quads example from the \texttt{html-mf-adr} subset, using the
W3C vCard vocabulary. Unlike Schema.org, no parent entity is linked, so the
place name is unavailable.}
\label{lst:vcard_nquads}
\end{listing}

\subsubsection{Address Filtering}
\label{sec:address-filtering}
To focus validation on records with sufficient spatial evidence, we retained only dirty records containing at least one core address signal: \textsf{streetAddress}, \textsf{addressLocality}, or \textsf{postalCode}.
This filtering step removes records whose address information is too sparse to be meaningfully validated, while preserving the vast majority of usable records.
After filtering, 219.9M dirty records remain from the Schema.org subsets (88.7\% retention rate) and 8.5M from the \textsf{html-mf-adr} subset (71.9\% retention rate).
Together, they form a combined pool of 228.4M dirty records.

\subsection{Ground Truth Generation}
\label{sec:data_validation}

To generate the ground truth, we need to find a true representation for each address.
This would normally require the definition of a desired target representation, followed by the manual verification of each dirty address.
Unfortunately, the latter is not scalable.
Alternatively, one can compare each dirty address with the corresponding representation in a trusted source.
The challenge is ensuring that each address is correctly matched.
Following this intuition, we resorted to a mixed approach, where we first tried to match each dirty address to its representation provided by a geocoding service, then applied manually crafted scripts to filter out inconsistencies.

\subsubsection{Geocoding Service Selection}
\label{sec:geocoding_selection}
Generating ground truth at this scale requires a geocoding service with broad geographic coverage, reliable address-level matches, and sufficient throughput for hundreds of millions of queries.
We therefore considered both proprietary geocoding services and open alternatives.
Proprietary services such as Google Places, HERE, and Geocode.xyz provide geocoding APIs, but their use at scale is constrained by a combination of licensing costs, rate limits, and regional coverage restrictions.
Some services are optimized for specific countries or regions, while others become impractical for 228.4M queries under their public rate limits or pricing models.
We also considered three open alternatives:
\emph{(i)}~Overture Maps\footnote{\url{https://overturemaps.org}}, a dataset of approximately 446M addresses stored as GeoParquet files on S3, which can be queried offline via DuckDB;
\emph{(ii)}~OpenAddresses\footnote{\url{https://openaddresses.io}}, a collection of roughly one billion addresses distributed as GeoJSON files, also queryable via DuckDB; and
\emph{(iii)}~Nominatim\footnote{\url{https://github.com/osm-search/Nominatim}}, the geocoding front-end of OpenStreetMap (OSM), which is available as a public RESTful API and can also be self-hosted to support high-throughput querying.

To assess their suitability for generating ground truth at scale, we compared them on a stratified sample of 1\emph{k} dirty addresses to preserve the original distribution of countries and points of interests (POIs).
For each service, we queried the sampled addresses and scored the returned matches based on fuzzy string similarity across house number, street name, postal code, and locality.
Nominatim was able to match with high confidence 96.5\% of the dirty addresses. We further verified a random sample of 1\emph{k}~addresses against the Google Address Validation API\footnote{\url{https://developers.google.com/maps/documentation/address-validation}}, finding 99.1\% agreement between the two sources.
The two alternatives showed substantial coverage gaps: Overture Maps and OpenAddresses covered only 39 and 19 countries, matching 10.8\% and 33.4\% of the dirty addresses, respectively.
Both significantly struggled with regions outside Europe.

The strong performance of Nominatim can be mainly attributed to two factors.
First, OSM maintains rich POI data, such as hotels, restaurants, and public amenities, that align well with the entities in our dataset.
Second, the structured search endpoint provided by Nominatim accepts individual address fields, which reduces ambiguity compared to free text.
Based on these findings, we selected Nominatim as the geocoding service for ground truth generation, and we deployed a self-hosted instance capable of handling several hundred requests per second.

\subsubsection{Geocoding Service Querying}
\label{sec:geocoding_querying}

\begin{listing}[t!]
\begin{lstlisting}[style=code,numbers=none,gobble=0]
   https://nominatim.openstreetmap.org/search
   .... ?amenity=extracted.name
   .... &street=extracted.streetAddress
   .... &city=extracted.addressLocality
   .... &postalcode=extracted.postalCode
   .... &state=extracted.addressRegion
   .... &country=extracted.addressCountry
   .... &format=geojson
   .... &limit=1
   .... &addressdetails=1
   .... &extratags=1
\end{lstlisting}

\vspace{-10pt}
\caption{The parameters used to query each dirty address through the Nominatim API.}
\label{fig:nominatim_search}
\end{listing}

For each dirty record, we queried the Nominatim API through its \textsf{/search} endpoint to retrieve the corresponding representation in the OSM database.
The query template, which exploits the structured information provided by the \textsf{PostalAddress} schema, is shown in Listing~\ref{fig:nominatim_search}.
Based on the additional parameters, Nominatim returns the closest match in JSON format. The result is a JSON object that includes address details and additional information about the location, such as opening hours.
Relevant fields for our use case are:
\emph{(i)}~\textsf{name}, i.e., the name associated
with the location;
\emph{(ii)}~\textsf{place\_rank}, an integer that reflects the granularity of the location (higher values correspond to more specific location types);
\emph{(iii)}~\textsf{address}, a nested JSON object that includes the address details of the location.
Unfortunately, the address details returned by Nominatim do not adhere to the PostalAddress schema.
We therefore need to perform schema alignment by defining appropriate mappings and transformations.

\subsubsection{Schema Alignment}
\label{sec:schema_alignment}

\begin{figure}[t!]
    \tikzset{>=latex}
    \centering
    \begin{tikzpicture}[
        node distance=5cm and 3cm,
        every node/.style={draw, text width=2.0cm,
        central/.style={text width=2.0cm, fill=yellow!20, font=\footnotesize},
        align=center, minimum height=0.5cm, font=\footnotesize},
        align=center, thick, scale=0.70
    ]

    \node[draw=none, text width=3cm] (col1) at (-4,10) {\textbf{Extracted Addresses} \\ (Before alignment)};
    \node[draw=none, text width=5cm] (col3) at (0,10) {\textbf{Aligned Schema} \\};
    \node[draw=none, text width=3cm] (col4) at (4,10) {\textbf{OSM Addresses} \\ (Before alignment)};

    \node (ex_name) at (-4,9) {name};
    \node (osm_name) at (4,9) {name};
    \node[central] (name) at (0,9) {name};
        \draw[->] (osm_name) -- (name);
        \draw[->] (ex_name) -- (name);

    \node (ext1) at (-4,7.5) {streetAddress};
    \node[central] (join1a) at (0,8) {road};
    \node[central] (join1b) at (0,7) {house\_number};
    \node (osm1a) at (4,8) {road};
    \node (osm1b) at (4,7) {house\_number};
        \draw[->] (ext1) -- (join1a);
        \draw[->] (ext1) -- (join1b);
        \draw[->] (osm1a) -- (join1a);
        \draw[->] (osm1b) -- (join1b);

    \node (ex_postcode) at (-4,6) {postalCode};
    \node (osm_postcode) at (4,6) {postcode};
    \node[central] (postcode) at (0,6) {postcode};
        \draw[->] (osm_postcode) -- (postcode);
        \draw[->] (ex_postcode) -- (postcode);

    \node (ex_local) at (-4,4) {addressLocality};
    \node[central] (local) at (0,4) {locality};
    \node (osm_local) at (4,4) {city, town,\\\textit{village}, \textit{borough}\\\textit{neighborhood}, \textit{hamlet},\\\textit{municipality}};
        \draw[->] (ex_local) -- (local);
        \draw[->] (osm_local) -- (local);

    \node (osm_region) at (4,1.5) {region, state\\\textit{county}, \textit{province}, \textit{state\_district}};
    \node (ex_region) at (-4,1.5) {addressRegion};
    \node[central] (region) at (0,1.5) {region};
        \draw[->] (osm_region) -- (region);
        \draw[->] (ex_region) -- (region);

    \node (ex_country) at (-4,-0.5) {addressCountry};
    \node[central] (countryA) at (0,-0) {country};
    \node[central] (countryB) at (0,-1) {country\_code};

    \node (osm_countryA) at (4,-0) {country};
    \node (osm_countryB) at (4,-1) {country\_code};
    \draw[->] (osm_countryA) -- (countryA);
    \draw[->] (osm_countryB) -- (countryB);
    \draw[->] (ex_country) -- (countryA);
    \draw[->] (ex_country) -- (countryB);

    \end{tikzpicture}
    \vspace{-.6cm}
    \caption{Schema alignment between extracted addresses (dirty records) and OSM addresses (ground truth).}
    \label{fig:schema_alignment}
\end{figure}
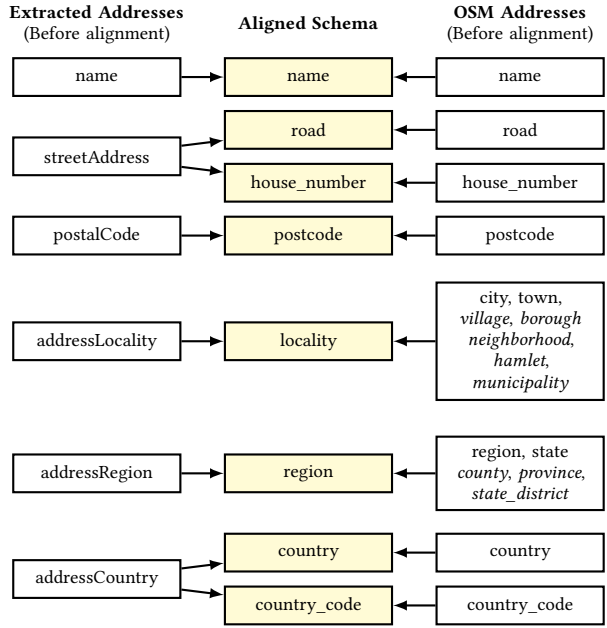

Figure~\ref{fig:schema_alignment} provides an overview of the designed alignment between the \textsf{PostalAddress} schema and the address details returned by Nominatim.
While \textsf{name} and \textsf{postcode} can be aligned directly, most fields require more refined strategies to handle structural differences.
For instance, OSM uses two independent fields to store the street name and the house number, which are instead combined into a single \textsf{streetAddress} attribute in \textsf{PostalAddress}.
To unify the two representations, we split \textsf{streetAddress} into \textsf{road} and \textsf{house\_number} to keep a more modular representation of the address.
We extracted the street name and house number from \textsf{streetAddress} using libpostal\footnote{\url{https://github.com/openvenues/libpostal}}, a popular address parser.
If libpostal could not identify one of the two components, the corresponding value was set to \textit{null} in the dirty version of the dataset.

Similarly, OSM stores both the country name and its ISO 3166-1 alpha-2 code\footnote{\url{https://www.iso.org/iso-3166-country-codes.html}}, each in a dedicated field.
In contrast, \textsf{PostalAddress} uses a single \textsf{addressCountry} attribute, which can contain either the country name or its code.
We decided to keep both fields in the aligned schema and to move the \textsf{addressCountry} value to \textsf{country\_code} if it matched with the required two-letter pattern.

Finally, \textsf{addressLocality} and \textsf{addressRegion} attributes have many possible correspondences in OSM, listed in Figure~\ref{fig:schema_alignment}.
Each location returned by Nominatim can contain one or more of these fields, requiring a flexible approach to identify the most appropriate value for the ground truth.
Specifically, we favored the \textsf{city} and \textsf{town} (\textsf{region} and \textsf{state}) fields, semantically closer to the concept of locality (region) in \textsf{PostalAddress}, considering the remaining ones only if none of them were present.
The edit distance from the extracted value, computed using the well-known Damerau-Levenshtein metric~\cite{DBLP:journals/cacm/Damerau64}, was then used as the tie-breaking criterion in case multiple (non)preferred fields were present at the same time.

\subsubsection{Ground Truth Filtering}
\label{sec:ground_truth_filtering}

After obtaining ground truth addresses from Nominatim, we applied a validation and normalization pipeline before constructing the benchmark dataset.

First, we filtered out 183.8M addresses (about 80\% of our queries) for which Nominatim did not return any matching location.
To further ensure completeness, we removed addresses whose corresponding ground truth entries were missing values for at least one attribute, excluding \textsf{name}.
Indeed, as \textsf{name} was missing in more than 42\% of the remaining entries, removing them would have significantly reduced the size of the dataset.
We opted therefore to create two versions of the dataset:
\emph{(i)}~\textbf{\dsnoname{}} (9,317,886 addresses), obtained by retaining all remaining addresses, dropping the \textsf{name} attribute from both the extracted dirty dataset and the ground truth;
\emph{(ii)}~\textbf{\dsname{}} (4,338,109 addresses), obtained from the portion of addresses for which the name was present in the ground truth, after a further refinement step using name similarity as a heuristic to identify and remove inaccurate matches.
We measured similarity through token set ratio\footnote{\url{https://rapidfuzz.github.io/RapidFuzz/Usage/fuzz.html\#token-set-ratio}} due to its robustness to different word ordering and partial overlaps.
A perfect threshold (100\%) was chosen after string normalization (lowercasing and converting to a canonical Unicode form) to ensure that distinct co-located entities were not mistakenly aligned.
In both dataset versions, duplicate addresses originating from the same source page and sharing identical attribute values in both the dirty dataset and the ground truth were removed.

Finally, to further improve the consistency of the \textsf{locality} attribute in the ground truth, we used publicly available country-based lists of recognized localities, e.g., Gemeinden in Germany\footnote{\url{https://public.opendatasoft.com/explore/assets/georef-germany-gemeinde/}} or places/subdivisions in the US\footnote{\url{https://public.opendatasoft.com/explore/assets/georef-united-states-of-america-place/} and \url{https://public.opendatasoft.com/explore/assets/georef-united-states-of-america-county-subdivision/}}, to maintain the same level of granularity in the \textsf{locality} attribute.
For each ground-truth entry, we checked the presence of a certified locality in that region from the list, allowing fuzzy matches based on containment or edit distance, and we updated the locality of that entry to the certified one in case it was reporting a different value.

\begin{table*}[t]
\centering
\footnotesize
\setlength{\tabcolsep}{8pt}
\caption{Number of distinct values for each attribute in the ground truth of the \dsname{} and \dsnoname{} datasets.}
\vspace{-.3cm}

\label{tab:distinct_values}
\begin{tabular}{lrrrrrrrrrr}
\toprule
Dataset version & \textsf{\_id} & \textsf{name} & \textsf{house\_number} & \textsf{road} & \textsf{postcode} & \textsf{locality} & \textsf{region} & \textsf{country} & \textsf{country\_code} & \textsf{graph\_iri}\\
\midrule
\dsname{} & 4,338,109 & 357,323 &  30,417 & 157,530  & 90,819  & 39,810 & 1,263& 128 & 127& 2,911,740\\
\dsnoname{} & 9,317,886  & --- & 74,539 & 412,231 &  152,099 & 61,579 & 1,461  & 135  &135 & 6,240,792\\
\bottomrule
\end{tabular}
\end{table*}

\subsection{Ethical Concerns}
\label{sec:ethical_concerns}

In alignment with prior work that analyzed the Common Crawl archives~\cite{subramani-etal-2023-detecting}, we refrained from including any explicit personally identifiable information (PII), such as SSNs, emails, phone numbers, and banking information.
After extraction, we also used the underlying PII detection library\footnote{\url{https://github.com/PovertyAction/PII_detection}} to identify unintentionally captured PII. No PII was detected.
The entities whose addresses we collected represent businesses and publicly known offices.
Furthermore, we semi-automatically audited the entries to identify facilities, further confirming that the addresses do not belong to individuals.

The data from Common Crawl is used under its Terms of Use.
In accordance with the Open Database Licence (ODbL) 1.0 of OpenStreetMap, our ground truth dataset remains under ODbL in line with the share-alike requirements.

\section{Dataset Statistics}
\label{sec:dataset-eval}

In this section, we report statistical insights from our analysis of the dataset in its different variations, i.e., \dsname{} and \dsnoname{}. As shown in Table~\ref{tab:distinct_values}, the schema of the generated dataset reflects the attributes defined in the schema alignment step (Section~\ref{sec:schema_alignment}): \textsf{house\_number}, \textsf{road}, \textsf{postcode}, \textsf{locality}, \textsf{region}, \textsf{country}, and \textsf{country\_code}.
The \textsf{name} attribute is present only in the \dsname{} version.
Further, a unique identifier (\textsf{\_id}) is assigned to every record in the whole collection to serve as its identifier. We also track the source page URL as an attribute (\textsf{graph\_iri}) for each record.

\begin{figure}[t]
  \centering
 \includegraphics[width=\columnwidth]{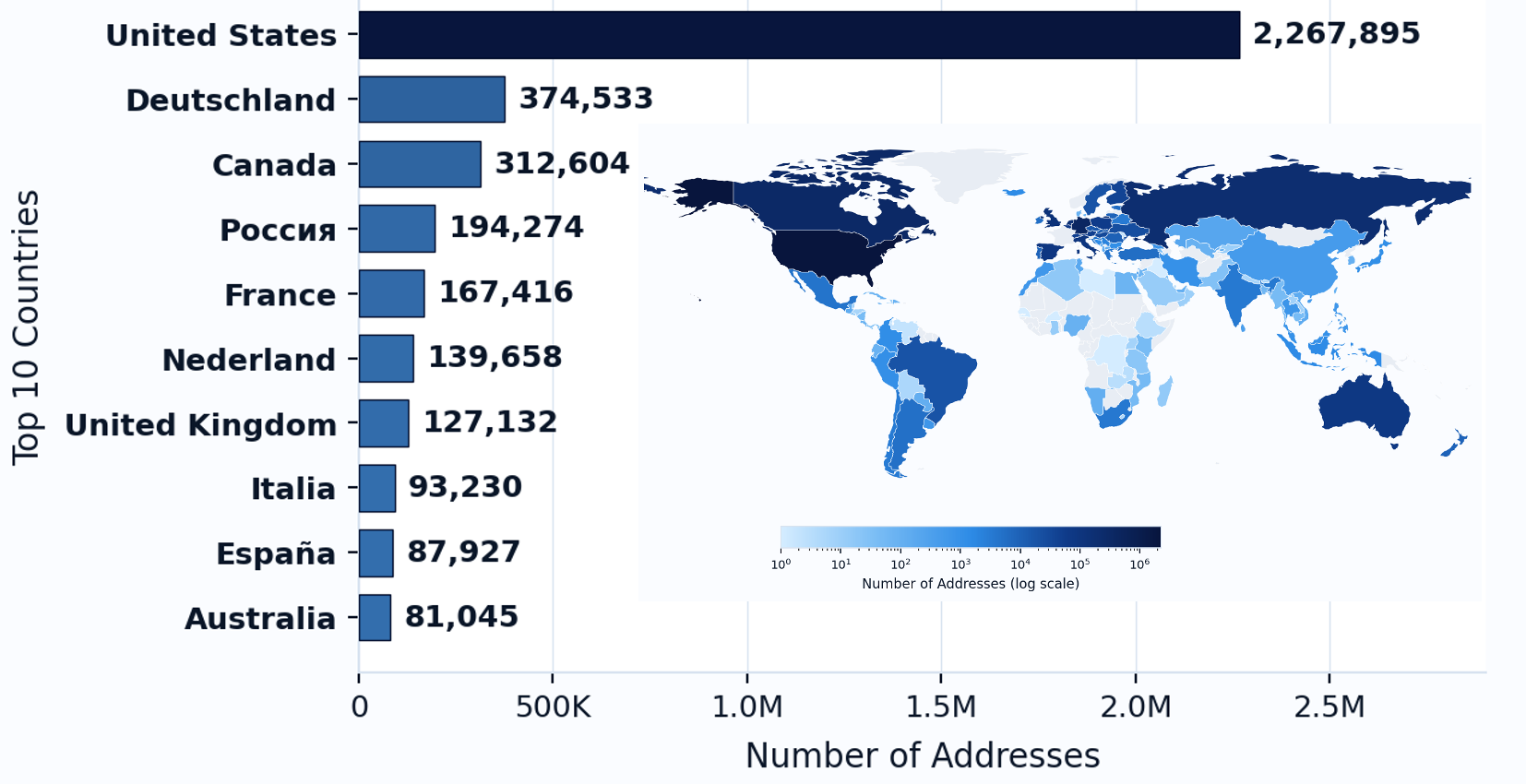}
 \vspace{-.6cm}

  \caption{Country distribution in the \dsname{} dataset.}
  \label{fig:geographic_distribution}
\end{figure}

\noindent\textbf{Geographic Distribution.}
Although country-level coverage is uneven, the dataset is geographically diverse, with addresses spanning 128 countries worldwide. Figure~\ref{fig:geographic_distribution} shows the geographic distribution of addresses in the \dsname{} version. If needed, one can readily sample a balanced subset over \emph{x} countries that remains larger than many existing cleaning datasets.

\noindent\textbf{Cardinalities.} Table~\ref{tab:distinct_values} reports the number of distinct values per attribute in the ground truth of the \dsname{} and \dsnoname{} versions.
With the only exception of the identifier, no attribute is unique.
At the record level, filtering out \textsf{\_id} and \textsf{graph\_iri} attributes, we can count 427,885 and 1,457,806 distinct records in the ground truth of the \dsname{} and \dsnoname{} versions, respectively.
In the \dsname{} version, multiple named entities can be located at the same address.
On average, each distinct address hosts 1.03 distinct entities, with a maximum of 51 co-located entities.

\noindent\textbf{Functional Dependencies.} Our dataset presents multiple functional dependencies (FDs), which can serve to detect errors and generate repairs in rule-based tools.
The \dsname{} version contains 14 FDs for \textsf{country} and 10 for \textsf{country\_code}, involving one (e.g., \textsf{country} $\rightarrow$ \textsf{country\_code}\footnote{Note that the rule \textsf{country\_code} $\rightarrow$ \textsf{country} does not hold instead, due to an exception, \textsf{Northern~Cyprus}, which is associated with the same ISO code as \textsf{Cyprus}.}), two (e.g., \textsf{postcode}, \textsf{locality} $\rightarrow$ \textsf{country}), or three attributes on the left-hand side (e.g., \textsf{house\_number}, \textsf{road}, \textsf{postcode} $\rightarrow$ \textsf{country}).
Country-based subsets may contain additional dependencies. For example, in German addresses \textsf{region} can be functionally determined by \textsf{postcode} and \textsf{locality}.
The full lists of FDs for the representative dataset variations described in the paper are publicly available in our GitHub repository.

\subsection{Dataset Slices}
\label{sec:dataset_variations}

For a finer-grained analysis, we readily sampled and made available multiple subsets of the \dsname{} dataset based on specific dimensions, such as the country and the type of point of interest (POI), to use them in our experimental evaluation.

\noindent\textbf{Country-based Slices.}
We first considered the geographic dimension, producing dedicated subsets for the two most represented countries (US and Germany).
As shown in Figure~\ref{fig:geographic_distribution}, the two slices (denoted as \textsf{us-slice} and \textsf{de-slice}) are composed of 2,267,895 and 374,533 addresses, respectively.

\noindent\textbf{POI-based Slices.}
We also created slices for 22 different types of POI, reflecting class-specific Schema.org subsets identified by WDC.
 This way, researchers can further explore specific cleaning strategies by augmenting certain entities with their type-specific properties~\cite{DBLP:journals/pvldb/LiC25, DBLP:journals/pvldb/LiuPCY17}.
In our analysis, we focus on \textsf{library} and \textsf{shopping-center}, which cover all identified error types.

\subsection{Error Statistics}
\label{subsec:err-rate-analysis}

To demonstrate the usefulness of our generated dataset, we analyze the characteristics of its errors. We report the overall error rate for the \dsname{} and \dsnoname{} variants and the error rate for individual address fields.
To capture not only frequency but also the nature of errors, we analyze the distribution of \textit{error types}.
In general, we compare the \emph{dirty} value in each cell with the corresponding \emph{clean} value from the ground truth. Cells that do not match exactly are counted as errors.

\begin{itemize}[leftmargin=*]
  \item \textbf{\dsname{}} -- Across 4,338,109 addresses with eight fields, hence 34,704,872 cells in total, \textbf{51.1\%} of the cells are clean, i.e., they match the ground truth, while \textbf{48.9\%} contain errors.
  \item \textbf{\dsnoname{}} -- Across 9,317,886 addresses with seven fields, hence 65,225,202 cells in total, \textbf{47.3\%} of the cells are clean, while \textbf{52.7\%} contain errors.
\end{itemize}

\subsubsection{Error Categories}
\label{subsec:err-type-analysis}

To characterize the errors present in our dataset, we apply a \emph{hierarchical, rule-based} classifier that assigns each cell to exactly one error category. Because a single cell can be subject to multiple errors simultaneously, e.g., a token reordering that also introduces a capitalization change, mutual exclusivity is enforced by a strict priority cascade: each rule is evaluated in the order listed below, and the first rule leading to a match with the ground truth determines the label. This design prevents double-counting and makes the classification reproducible, ensuring that aggregate statistics are fully comparable across dataset slices.

Before any rule is applied, both the dirty and clean value are normalized deterministically: zero-width and non-breaking space characters are removed or replaced with regular spaces, whitespace runs are collapsed, and Unicode NFC normalization is applied. This step is intentionally case- and punctuation-preserving so that purely cosmetic differences are absorbed into the \emph{Clean} category rather than triggering downstream rules.

The taxonomy comprises 12 mutually exclusive categories, applied in the following priority order:

\smallskip

\noindent
\emph{Clean} $\to$
\emph{Missing} $\to$
\emph{Parsing} $\to$
\emph{Capitalization} $\to$
\emph{Special Chars} $\to$
\emph{Name Reformat} $\to$
\emph{Language} $\to$
\emph{Abbreviation} $\to$
\emph{Token Subset} $\to$
\emph{Typo} $\to$
\emph{Word Substitution} $\to$
\emph{Miscellaneous}

\smallskip

\noindent
We begin by separating exact matches and missing values, which can be identified without applying any transformation. Next come deterministic representation changes such as parsing issues, capitalization, punctuation, and name reformatting. These must be resolved early to avoid misclassifying them into broader, less precise categories such as typos. Language variations and abbreviations, which produce values that are lexically distinct but semantically equivalent, are evaluated before token-subset and edit-distance-based typo rules. Finally, typos, word substitutions, and miscellanea capture progressively less constrained differences, serving as a catch-all for cases not resolved by the earlier, more specific rules.
Next, we explain each category in detail.

\begin{enumerate}[leftmargin=*]

    \item \textbf{Clean} -- The normalized dirty value is identical to the normalized ground truth. No error is present.

    \item \textbf{Missing} -- The dirty value is empty after normalization. This error type is placed first, as applying any subsequent rule to an empty string would be meaningless.

    \item \textbf{Parsing} -- The dirty value contains an artifact resulting from an incorrectly identified character encoding, e.g., \textsf{M\%C3\%BCnchen} $\leftrightarrow$ \textsf{München}, \textsf{Caf\&\#233;} $\leftrightarrow$ \textsf{Café}.
    Equality with the ground truth is restored after an appropriate decoding step.

    \item \textbf{Capitalization} -- The two values are identical under Unicode case folding~\cite{unicode2024}, e.g., \textsf{ATLANTA} $\leftrightarrow$ \textsf{Atlanta}, \textsf{10a} $\leftrightarrow$ \textsf{10A}.

    \item \textbf{Special Characters} -- The two values are identical under case folding after removing all Unicode punctuation, symbol, and whitespace characters from both, e.g., \textsf{Notre-Dame Street} $\leftrightarrow$ \textsf{Notre Dame Street}, \textsf{U.S.A.} $\leftrightarrow$ \textsf{USA}.
    This rule fires for differences that are entirely attributable to punctuation usage conventions.

    \item \textbf{Name Reformat} -- The multiset of accent-folded, lowercased word-level tokens is identical for both values, despite different ordering in the original strings, e.g., \textsf{Howson Branch, Austin Public Library} $\leftrightarrow$ \textsf{Austin Public Library -- Howson Branch}, \textsf{South 37th Avenue} $\leftrightarrow$ \textsf{37th Avenue South}.

    \item \textbf{Language} -- The two values refer to the same entity but differ due to a language variation, such as a translation, an exonym, or a romanisation, e.g., \textsf{Austria} $\leftrightarrow$ \textsf{\"{O}sterreich}, \textsf{Athens} $\leftrightarrow$ \textsf{$A\theta\acute{\eta}\nu\alpha$}.
    Resolving language variations before abbreviations prevents short tokens in a non-Latin script from being claimed by the abbreviation lookup, which operates on Latin-script expansions.

    \item \textbf{Abbreviation} -- The two values are equal after standardized address-form expansion via libpostal~\cite{barrington2017libpostal}, e.g., \textsf{Ave} $\leftrightarrow$ \textsf{Avenue}, \textsf{Bachmannstr.} $\leftrightarrow$ \textsf{Bachmannstraße}.

    \item \textbf{Token Subset} -- One token multiset is a strict subset of the other, e.g., \textsf{New York} $\subset$ \textsf{New York City}, \textsf{Finavia} $\subset$ \textsf{Finavia Oyj}. Structural containment is a stronger indicator than a small edit distance: resolving such cases before typos prevents us from incorrectly labeling them as typographical errors.

    \item \textbf{Typo} -- The two normalized strings differ by at most two edit operations (substitution, insertion, deletion, or adjacent transposition) under the Damerau--Levenshtein distance~\cite{levenshtein1966binary,DBLP:journals/cacm/Damerau64}, e.g., \textsf{Theater} $\leftrightarrow$ \textsf{Theatre}, \textsf{Marriott} $\leftrightarrow$ \textsf{Mariot}.

    \item \textbf{Word Substitution} -- Exactly one token differs between the two multisets while at least one token is shared, e.g., \textsf{VA Clinic} $\leftrightarrow$ \textsf{Outpatient Clinic}. This category captures factual inaccuracies, but also morphological variants, derivational forms, and near-synonyms that represent the same concept.

    \item \textbf{Miscellaneous} -- Any cell-level difference that does not satisfy any of the preceding rules is assigned this residual category. It subsumes cases such as cross-source conflicts, e.g., \textsf{California} $\leftrightarrow$ \textsf{Highway 76}, administrative reassignments, e.g., \textsf{Strathroy} $\leftrightarrow$ \textsf{Adelaide Metcalfe}, OCR-style corruptions, etc.

\end{enumerate}

Our categories build on prior work on error detection and taxonomies, but are more fine-granular and tailored to cell-level differences. Bhadauria et al.~\cite{DBLP:journals/corr/abs-2604-09277} provide a broader catalog of errors. Our taxonomy covers a subset observable through cell-wise address comparison and focuses on the cause of the error and not the type they belong to based on the algorithm that exposes them, such as rule violation or outliers~\cite{AbedjanCDFIOPST16}.

\begin{table*}[t]
\centering
\scriptsize
\setlength{\tabcolsep}{4pt}
\caption{Distribution of error types across dataset variations (percentages of total cells). \emph{Parsing} is also reported as an absolute count due to its negligible overall frequency. Statistics for other POI-based slices are available on GitHub.}
\vspace{-.3cm}

\label{tab:facility-error-summary}
\resizebox{\textwidth}{!}{
\begin{tabular}{lrrrrrrrrrrrrrrr}
\toprule
Slice & Num-rows & Clean & Missing & Parsing (\#) & Parsing (\%) & Capitalization & Special Chars & Name Reformat & Language & Abbreviation & Token Subset & Typo & Word Subst. & Misc \\
\midrule
\dsname{} & 4{,}338{,}109 & 51.1\% & 23.7\% & 8830 & 0.025\% & 2.1\% & 0.8\% & 0.6\% & 6.9\% & 9.3\% & 2.6\% & 1.2\% & 0.2\% & 1.5\% \\
\dsnoname{} & 9{,}317{,}886 & 47.3\% & 28.6\% & 500 & 0.001\% & 1.7\% & 0.8\% & 0.3\% & 6.4\% & 9.5\% & 1.8\% & 1.3\% & 0.4\% & 1.9\% \\
\midrule
\usslice{} & 2{,}267{,}895 & 52.9\% & 20.1\% & 6940 & 0.038\% & 1.0\% & 0.5\% & 0.1\% & 7.0\% & 14.9\% & 2.2\% & 0.7\% & 0.0\% & 0.5\% \\
\deslice{} & 374{,}533 & 53.5\% & 30.0\% & 91 & 0.003\% & 1.9\% & 1.2\% & 0.3\% & 5.3\% & 1.2\% & 2.7\% & 0.3\% & 0.0\% & 3.6\% \\
\ruslice{} & 194,274 & 51.9\% & 29.6\% & 0 & 0.000\% & 7.8\% & 0.3\% & 0.1\% & 2.2\% & 0.4\% & 0.7\% & 6.4\% & 0.2\% & 0.3\% \\

\midrule
\library{} & 5{,}370 & 51.7\% & 27.3\% & 10 & 0.023\% & 2.8\% & 0.4\% & 0.6\% & 2.0\% & 5.8\% & 1.6\% & 1.1\% & 0.0\% & 6.6\% \\
\shoppingcenter{} & 3{,}478 & 45.8\% & 22.4\% & 13 & 0.047\% & 3.1\% & 0.2\% & 7.3\% & 4.2\% & 7.1\% & 2.5\% & 6.5\% & 0.2\% & 0.6\% \\
\bottomrule
\end{tabular}}
\end{table*}

\begin{figure}[t]
  \centering
  \includegraphics[width=\columnwidth]{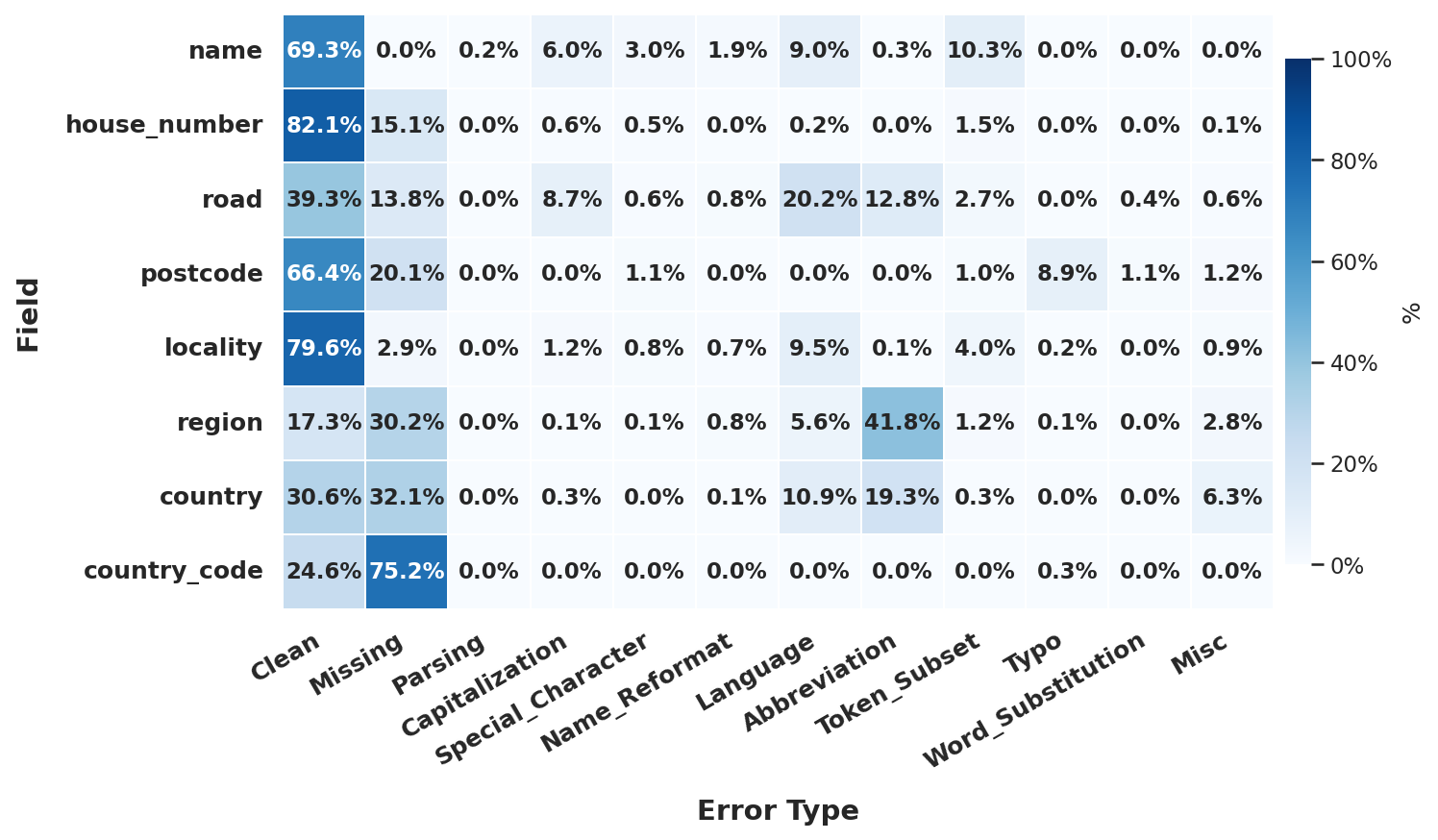}
  \vspace{-.8cm}
  \caption{Percentages of clean and erroneous cells (grouped by error type) per attribute in the \dsname{} dataset.}\label{fig:full-heatmap}

\end{figure}

\subsubsection{Error Category Distributions}

We now report the ratios of the defined error types for the full dataset and its slices.

\subsubsection*{Full Dataset}
Figure~\ref{fig:full-heatmap} summarizes the cell-wise comparison over all attributes between the extracted dataset and its ground truth for the \dsname{} version. The \dsnoname{} version shows a very similar distribution. The largest drift between the \emph{dirty} and \emph{clean} version of the dataset is due to representation issues and missing values. Common issues include abbreviated or shortened forms of street and region names, partial or omitted locality fields, and inconsistent capitalization or punctuation. Textual attributes such as \textsf{region} and \textsf{country} are the most affected, often due to incomplete or shortened representations of organizations and administrative divisions. In contrast, structured fields, such as \textsf{house\_number}, \textsf{postcode}, and \textsf{locality}, tend to be considerably cleaner, with most inconsistencies limited to minor formatting or typographic deviations. The \textsf{country\_code} value is frequently missing, reflecting the common omission of standardized codes on Web sources.

\subsubsection*{Country-based Slices}
The distribution of error types in addresses from \usslice{} is very similar to the one in the full dataset, where US addresses represent more than the half of the entries.
US addresses show a significantly higher rate of abbreviations, especially in the \textsf{region} attribute (74.6\%), but also for \textsf{road} (25.3\%) and \textsf{country} (19\%). For instance, addresses inconsistently use abbreviations for street components, such as ``W'' for ``West'' or ``Blvd.'' for ``Boulevard''.
The abbreviation rate is much lower in \deslice{}, e.g., 9.2\% for \textsf{road} or close to 0\% for \textsf{region} and \textsf{country}, which shows instead a high rate of missing values for \textsf{region} (70.3\%). Indeed, for \emph{free cities}, such as Berlin and Hamburg, the region coincides with the city itself and is therefore omitted for practical reasons. Another distinguishing factor is the overall clean ratio for street names (61.8\%). An interesting observation concerns the orthography of the word ``street'' (``Stra{\ss}e'') in German and Swiss addresses. While German addresses strictly adhere to the usage of ``\ss'' instead of ``ss'' in the ground truth, Swiss addresses do the opposite. The correspondingly wrong usage can be observed in the dirty version of the dataset.

\subsubsection*{POI-based Slices}

In Table~\ref{tab:facility-error-summary}, we report the distribution of error types for the two POI-based slices (\textsf{library} and \textsf{shopping-center}) used in our experiments.
Statistics for the other 20 slices are available on GitHub.
Missing values are by far the most common error type across all slices, ranging from 16\% to 49.1\% of the cells, with the highest rates concentrated in sport-related slices.
Abbreviations are dominating among representational inconsistencies (median 9.2\%, up to 13.8\%), while typos are rare in most slices (median 0.8\%, up to 6.9\%).
The remaining categories vary substantially by POI type.
Language variations range from 2\% to 16.2\% across slices, reaching the highest values in tourism-related slices, such as \textsf{ski-resort}, often referenced from multiple countries.
Focusing on the two selected slices, \textsf{library} combines a high rate of missing values (27.3\%) with the largest amount of miscellaneous errors across all slices (6.6\%), while \textsf{shopping-center} exhibits the highest rate of name reformatting (7.3\%) and the second-highest rate of typos (6.5\%), together with a lower rate of clean cells (45.8\% vs.\ 51.7\%).

\section{Experimental Analysis}\label{sec:experiments}
To demonstrate the utility of our benchmark dataset, we conducted an extensive set of experiments and evaluated the effectiveness and scalability of state-of-the-art data cleaning systems across different subsets. We selected representative systems for rule-based, learning-based, and LLM-driven approaches, and assessed their performance in terms of effectiveness, runtime, and memory usage.

\subsubsection*{Setup}\label{sec:setup}
All dataset variations and source code for dataset extraction, ground truth generation, analysis, and experiments are publicly available on
GitHub.
The code is implemented in Python. Runtime experiments were executed on a server equipped with an AMD EPYC 7702P 64-core processor and 512~GB of RAM.
RAID-configured SSDs were used for storage.

\subsection{Datasets}
To evaluate state-of-the-art cleaning systems across different scenarios, we created the samples summarized in Table~\ref{tab:dataset_slices}.
For effectiveness experiments, we used samples of 10\emph{k} records of \dsname{}, \deslice{}, and \ruslice{}.
The latter contains Russian addresses and was selected to assess how cleaning systems perform on a non-Latin address subset.
Since US addresses already represent the majority in \dsname{}, we exclude the results on \usslice{} for brevity.
As the error rate in the original dataset is rather high (about 50\% of the cells), we also generated samples at lower error rates by proportionally replacing erroneous cells with the corresponding ground truth values.
For example, on a dataset with 100 cells among which 50 are marked as erroneous, hence presenting a 50\% error rate, a 20\% error rate would be obtained by randomly restoring 30 of the 50 erroneous cells to their ground truth values.
We additionally report on two POI-based slices, \library{} and \shoppingcenter{}, which differ with regard to their error-type distributions, to assess system behavior on domain-specific data.

For scalability experiments, we considered random \dsname{} samples of 10\emph{k}, 100\emph{k}, 500\emph{k}, and 1M records.
All datasets have eight columns: we excluded \textsf{\_id} and \textsf{graph\_iri} from all experiments, as both are always clean.
For the effectiveness experiments, we additionally filtered out records that were exact duplicates in their dirty version after the removal of the \textsf{\_id} and \textsf{graph\_iri} columns.
This determines the error rates marked as original* in Table~\ref{tab:dataset_slices}, which are slightly different from those reported in Table~\ref{tab:facility-error-summary}.

\begin{table}[t!]
\centering
\scriptsize
\caption{Overview of dataset samples used in the experiments. For the effectiveness experiments, records that were exact duplicates in their dirty version after the removal of the \textsf{\_id} and \textsf{graph\_iri} columns were filtered out before sampling.}
\vspace{-.2cm}
\label{tab:dataset_slices}
\renewcommand{\arraystretch}{1.1}
\begin{tabular}{lccr}
\hline
\textbf{Dataset} & \textbf{Error Rate} & \textbf{\#Rows} & \textbf{\#Columns} \\
\hline
\multicolumn{4}{l}{\textit{Effectiveness Experiments}} \\
\hline
\dsname{} & 0.05, 0.10, 0.20, 0.51 (original*) & 10,000 & 8 \\
\deslice{} & 0.05, 0.10, 0.20, 0.48 (original*) & 10,000 & 8 \\
\ruslice{} & 0.05, 0.10, 0.20, 0.65 (original*) & 10,000 & 8 \\
\library{} & 0.64 (original*) & 1,656 & 8 \\
\shoppingcenter{} & 0.64 (original*) & 802 & 8 \\
\hline
\multicolumn{4}{l}{\textit{Scalability Experiments}} \\
\hline
\dsname{} (All) & 0.49 & 4,338,109 & 8 \\
\dsname{} (1M) & 0.49 & 1,000,000 & 8 \\
\dsname{} (500$k$) & 0.49 & 500,000 & 8 \\
\dsname{} (100$k$) & 0.49 & 100,000 & 8 \\
\dsname{} (10$k$) & 0.49 & 10,000 & 8 \\
\hline
\end{tabular}
\end{table}

\subsection{Baselines}
In our experiments, we considered several state-of-the-art data cleaning systems
in representation of rule-based, learning-based, and LLM-driven approaches.

\subsubsection*{Rule-based Approaches}
\hfill\\
\noindent{\textbf{Horizon}}~\cite{DBLP:journals/pvldb/RezigOAEMS21} is a scalable rule-based system that generates repairs in linear time. We used the implementation provided by the authors\footnote{\url{https://github.com/D2IP-TUB/Horizon}}. As input, Horizon requires a set of FDs: we extracted them from the clean version of each sample using TANE~\cite{DBLP:journals/cj/HuhtalaKPT99}, relying on a third-party Python implementation\footnote{\url{https://github.com/codocedo/tane}}.

\noindent{\textbf{UniClean}}~\cite{DBLP:journals/pvldb/DingQWCTSHW25} is a system for building efficient data cleaning workflows. We classify it as rule-based because it requires predefined rules and constraints as input. We used the official implementation\footnote{\url{https://github.com/qzkinhit/UniClean-bench-Result}} with its default configuration and the same set of FDs used for Horizon. In addition, we enabled pattern-based and missing-value cleaners, as UniClean supports multiple error types.
For datasets where UniClean did not finish within the allocated time budget of 10 hours under its initial configuration, we reduced the workload by incrementally modifying three settings. First, we changed the sampling mode, choosing whether UniClean derives the repair plan from a quality-based coreset or from the full data directly. Second, we simplified the set of FDs by removing those with three or more attributes on the left-hand side. Third, we adjusted the block-size threshold \textsf{single\_max}. When repairing a target attribute or attribute group, UniClean applies all rules jointly if the sampled block contains fewer than \textsf{single\_max} rows; otherwise, it applies the rules one at a time.
For each dataset we applied these changes incrementally in the presented order, stopping as soon as the run completed within the time budget. The final configuration used for each dataset is available in our repository.
\subsubsection*{Learning-based Approaches}
\hfill\\
\noindent{\textbf{HoloClean}}~\cite{DBLP:journals/pvldb/RekatsinasCIR17} is an error correction system that combines integrity constraints, external data sources, and statistical profiles in a unified factor graph model to infer the most likely repairs for detected errors. We used the original implementation\footnote{\url{https://github.com/HoloClean/holoclean}} and provided it with the same FDs used for Horizon and UniClean.

\noindent{\textbf{Raha}}~\cite{DBLP:conf/sigmod/MahdaviAFMOS019} and \textbf{Baran}~\cite{DBLP:journals/pvldb/MahdaviA20}
are semi-supervised systems for error detection and correction, respectively. Both formulate detection and correction as classification tasks and learn data-specific patterns to identify and repair erroneous cells. All experiments were conducted using their official implementation\footnote{\url{https://github.com/BigDaMa/raha}}. As recommended by the authors, we ran each experiment with 20 labeled tuples per task to train the corresponding models.

\noindent{\textbf{DataWig}}~\cite{DBLP:journals/jmlr/BiessmannRSNSTL19} is a data imputation method based on deep learning. We used the official implementation\footnote{\url{https://github.com/awslabs/datawig}} and the SimpleImputer module for our experiments. We used 20 labeled tuples for training and validation. For each column, the imputer was trained to predict its values based on all remaining columns. We then used the calibrated probability scores provided by the imputer to filter predictions, setting the threshold to 0.5. Among the remaining predictions, those differing from the original values were considered as detected errors. For the correction task, we then used the trained models to predict repairs for all actual erroneous cells in the dataset.

\subsubsection*{LLM-driven Approaches}
\hfill\\
\noindent{\textbf{Glock's}}~\cite{glock2150detecting} recently used LLMs to detect and correct errors in personal contact information datasets, a use case with which our dataset closely aligns. The proposed approach prompts LLMs to process each record individually, providing table-level rules and descriptions to guide both detection and correction. We used the official implementation\footnote{\url{https://github.com/Anna-Christina-Glock/pci-llm-toolkit}} and adapted the prompts to fit our dataset, making the edited versions publicly available in our repository. Experiments were conducted using Gemini 2.5 Flash-Lite. In preliminary tests, we had observed very similar results between Gemini 2.5 Flash-Lite and GPT-5-mini. We opted for Gemini 2.5 Flash-Lite due to its faster response times.

\begin{table*}[t!]
\scriptsize
\centering
\caption{Error detection and correction results across POI-based slices.}
\vspace{-.3cm}

\label{tab:facility_results}
\renewcommand{\arraystretch}{1.1}
\begin{tabular}{l|lccc|lccc}
\hline
\textbf{Dataset} & \textbf{Error Detection System} & \textbf{Precision} & \textbf{Recall} & \textbf{F1-Score} & \textbf{Error Correction System} & \textbf{Precision} & \textbf{Recall} & \textbf{F1-Score} \\
\hline
\multirow{6}{*}{\library{}}
& Raha & 0.92  & 0.91 &  0.91 & Baran & 0.14 & 0.11 & 0.12 \\
& Horizon & 0.49 & 0.54 &  0.51 & Horizon & 0.01 & 0.005 & 0.007 \\
& DataWig & 0.66 & 0.97 & 0.79 & DataWig &0.30 & 0.29 &0.29  \\
& HoloClean &  0.88 & 0.73 & 0.80 & HoloClean & 0.001 & 0.001 & 0.001\\
& Glock's &0.97 & 0.56 &0.71 & Glock's & 0.31 & 0.27 & 0.29\\
& UniClean & 0.73 & 0.86 & 0.79 & UniClean & 0.005 & 0.03 & 0.009 \\
\hline
\multirow{6}{*}{\shoppingcenter{}}
& Raha & 0.90 &  0.90 & 0.90  & Baran & 0.47 & 0.35 & 0.40  \\
& Horizon & 0.54  & 0.65 & 0.59 & Horizon & 0.004 & 0.003 & 0.003 \\
& DataWig & 0.67 &  0.98 & 0.80 & DataWig & 0.28  & 0.27 &  0.27 \\
& HoloClean & 0.91 & 0.70 & 0.79 & HoloClean & 0.01 & 0.003 & 0.004 \\
& Glock's & 0.98 & 0.60 & 0.74 & Glock's & 0.27  & 0.24& 0.25\\
& UniClean & 0.95 & 0.67 & 0.78 & UniClean & 0.01 & 0.02 & 0.01 \\
\hline
\end{tabular}
\end{table*}

\subsection{Effectiveness Experiments}

\begin{figure*}[t!]
  \centering
  \begin{tikzpicture}
    \begin{groupplot}[
      group style={
        group size=6 by 1,
        horizontal sep=1cm,
      },
      width=0.2\textwidth,
        height=0.2\textwidth,
      xlabel={Error Rate},
      ylabel={F1-Score},
      xmode=log,
      xmin=0, xmax=0.55,
      ymin=0, ymax=1,
      xtick={0.05, 0.10, 0.20, 0.52},
       tick label style={font=\normalsize, align=center, yshift=-0.5ex},
      xticklabels={$5\%$, $ $, $20\%$, $51\%$},
      grid=both,
    ]

    \nextgroupplot[
      title={\deslice{} - Detection},
      legend to name=sharedlegend,
      legend columns=-1,
      legend style={draw=none, fill=none},
      xmin=0, xmax=0.55,
      ymin=0, ymax=1,
      xtick={0.05, 0.10, 0.20, 0.48},
       tick label style={font=\normalsize, align=center, yshift=-0.5ex},
      xticklabels={$5\%$, $ $, $20\%$, $48\%$},
    ]
    \addplot[
      color=blue,
      mark=o,
    ] table[x=error_rate, y=det_f1, col sep=comma]
    {experiments_results/effectiveness/de_slice/raha.csv};\addlegendentry{Raha}
    \addplot[
      color=Emerald,
      mark=diamond,
    ] table[x=error_rate, y=det_f1, col sep=comma]
    {experiments_results/effectiveness/de_slice/datawig.csv};\addlegendentry{DataWig}
    \addplot[
      color=purple,
      mark=star,
    ] table[x=error_rate, y=det_f1, col sep=comma]
    {experiments_results/effectiveness/de_slice/glocks.csv};\addlegendentry{Glock's}
    \addplot[
      color=Brown,
      mark=oplus,
    ] table[x=error_rate, y=det_f1, col sep=comma]
    {experiments_results/effectiveness/de_slice/holoclean.csv};\addlegendentry{HoloClean}
     \addplot[
      color=Plum,
      mark=square*,
    ] table[x=error_rate, y=det_f1, col sep=comma]
    {experiments_results/effectiveness/de_slice/horizon.csv};\addlegendentry{Horizon}

    \addplot[
      color=WildStrawberry,
      mark=pentagon,
    ] table[x=error_rate, y=det_f1, col sep=comma]
    {experiments_results/effectiveness/de_slice/uniclean.csv};
    \addlegendentry{UniClean}

    \addlegendimage{color=BurntOrange, mark=*}\addlegendentry{Baran}

    \nextgroupplot[
      title={\ruslice{} - Detection},
      ylabel={},
      xmin=0, xmax = 0.72,
       xtick={0.05, 0.10, 0.20, 0.65},
      xticklabels={$5\%$, $ $, $20\%$, $65\%$},
    ]
    \addplot[
      color=blue,
      mark=o,
    ] table[x=error_rate, y=det_f1, col sep=comma]
    {experiments_results/effectiveness/russia/raha.csv};
    \addplot[
      color=Emerald,
      mark=diamond,
    ] table[x=error_rate, y=det_f1, col sep=comma]
    {experiments_results/effectiveness/russia/datawig.csv};
    \addplot[
      color=purple,
      mark=star,
    ] table[x=error_rate, y=det_f1, col sep=comma]
    {experiments_results/effectiveness/russia/glocks.csv};
    \addplot[
      color=Brown,
      mark=oplus,
    ] table[x=error_rate, y=det_f1, col sep=comma]
    {experiments_results/effectiveness/russia/holoclean.csv};
     \addplot[
      color=Plum,
      mark=square*,
    ] table[x=error_rate, y=det_f1, col sep=comma]
    {experiments_results/effectiveness/russia/horizon.csv};

    \addplot[
      color=WildStrawberry,
      mark=pentagon,
    ] table[x=error_rate, y=det_f1, col sep=comma]
    {experiments_results/effectiveness/russia/uniclean.csv};

    \nextgroupplot[
      title={\textsf{full} - Detection},
      ylabel={}
    ]
    \addplot[
      color=blue,
      mark=o,
    ] table[x=error_rate, y=det_f1, col sep=comma]
    {experiments_results/effectiveness/full/raha.csv};
    \addplot[
      color=Emerald,
      mark=diamond,
    ] table[x=error_rate, y=det_f1, col sep=comma]
    {experiments_results/effectiveness/full/datawig.csv};
    \addplot[
      color=purple,
      mark=star,
    ] table[x=error_rate, y=det_f1, col sep=comma]
    {experiments_results/effectiveness/full/glocks.csv};
    \addplot[
      color=Brown,
      mark=oplus,
    ] table[x=error_rate, y=det_f1, col sep=comma]
    {experiments_results/effectiveness/full/holoclean.csv};
     \addplot[
      color=Plum,
      mark=square*,
    ] table[x=error_rate, y=det_f1, col sep=comma]{experiments_results/effectiveness/full/horizon.csv};

    \addplot[
      color=WildStrawberry,
      mark=pentagon,
    ] table[x=error_rate, y=det_f1, col sep=comma]
    {experiments_results/effectiveness/full/uniclean.csv};

    \nextgroupplot[
      title={\deslice{} - Correction},
      ylabel={},
      xmin=0, xmax=0.55,
      ymin=0, ymax=1,
      xtick={0.05, 0.10, 0.20, 0.48},
       tick label style={font=\normalsize, align=center, yshift=-0.5ex},
      xticklabels={$5\%$, $ $, $20\%$, $48\%$},
    ]

    \addplot[
      color=Emerald,
      mark=diamond,
    ] table[x=error_rate, y=corr_f1, col sep=comma]
    {experiments_results/effectiveness/de_slice/datawig.csv};

    \addplot[
      color=purple,
      mark=star,
    ] table[x=error_rate, y=corr_f1, col sep=comma]
    {experiments_results/effectiveness/de_slice/glocks.csv};

    \addplot[
      color=Brown,
      mark=oplus,
    ] table[x=error_rate, y=corr_f1, col sep=comma]
    {experiments_results/effectiveness/de_slice/holoclean.csv};

    \addplot[
      color=Plum,
      mark=square*,
    ] table[x=error_rate, y=corr_f1, col sep=comma]
    {experiments_results/effectiveness/de_slice/horizon.csv};

    \addplot[
      color=BurntOrange,
      mark=*,
    ] table[x=error_rate, y=corr_f1, col sep=comma]
    {experiments_results/effectiveness/de_slice/baran.csv};

    \addplot[
      color=WildStrawberry,
      mark=pentagon,
    ] table[x=error_rate, y=corr_f1, col sep=comma]
    {experiments_results/effectiveness/de_slice/uniclean.csv};

    \nextgroupplot[
      title={\ruslice{} - Correction},
      ylabel={},
      xmin=0, xmax = 0.72,
       xtick={0.05, 0.10, 0.20, 0.65},
      xticklabels={$5\%$, $ $, $20\%$, $65\%$},
    ]

    \addplot[
      color=Emerald,
      mark=diamond,
    ] table[x=error_rate, y=corr_f1, col sep=comma]
    {experiments_results/effectiveness/russia/datawig.csv};

    \addplot[
      color=purple,
      mark=star,
    ] table[x=error_rate, y=corr_f1, col sep=comma]
    {experiments_results/effectiveness/russia/glocks.csv};

    \addplot[
      color=Brown,
      mark=oplus,
    ] table[x=error_rate, y=corr_f1, col sep=comma]
    {experiments_results/effectiveness/russia/holoclean.csv};

    \addplot[
      color=Plum,
      mark=square*,
    ] table[x=error_rate, y=corr_f1, col sep=comma]
    {experiments_results/effectiveness/russia/horizon.csv};

    \addplot[
      color=BurntOrange,
      mark=*,
    ] table[x=error_rate, y=corr_f1, col sep=comma]
    {experiments_results/effectiveness/russia/baran.csv};

    \addplot[
      color=WildStrawberry,
      mark=pentagon,
    ] table[x=error_rate, y=corr_f1, col sep=comma]
    {experiments_results/effectiveness/russia/uniclean.csv};

    \nextgroupplot[
      title={\textsf{full} - Correction},
      ylabel={}
    ]

    \addplot[
      color=Emerald,
      mark=diamond,
    ] table[x=error_rate, y=corr_f1, col sep=comma]
    {experiments_results/effectiveness/full/datawig.csv};
    \addplot[
      color=Brown,
      mark=oplus,
    ] table[x=error_rate, y=corr_f1, col sep=comma]
    {experiments_results/effectiveness/full/holoclean.csv};
    \addplot[
      color=purple,
      mark=star,
    ] table[x=error_rate, y=corr_f1, col sep=comma]
    {experiments_results/effectiveness/full/glocks.csv};
    \addplot[
      color=Plum,
      mark=square*,
    ] table[x=error_rate, y=corr_f1, col sep=comma]
    {experiments_results/effectiveness/full/horizon.csv};

\addplot[
      color=WildStrawberry,
      mark=pentagon,
    ] table[x=error_rate, y=corr_f1, col sep=comma]
    {experiments_results/effectiveness/full/uniclean.csv};

\addplot[
      color=BurntOrange,
      mark=*,
    ] table[x=error_rate, y=corr_f1, col sep=comma]
    {experiments_results/effectiveness/full/baran.csv};

    \end{groupplot}

    \node at ($(group c3r1.south)!0.5!(group c4r1.south)+(0,-1.2cm)$)
      {\ref{sharedlegend}};

  \end{tikzpicture}

\vspace{-.5cm}

  \caption{F1-score of different data cleaning systems for error detection and correction on subsets of \deslice{}, \ruslice{}, and \dsname{} dataset variations at different error rates.}
  \label{fig:effectiveness}
\end{figure*}
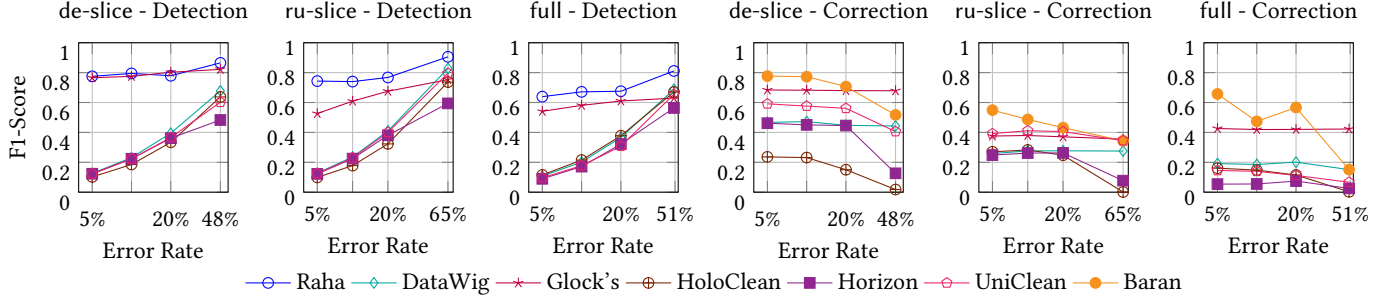

We evaluate the effectiveness of all baselines for both error detection and correction across different error rates, country slices, and POI-based slices. We ran each system three times and report the average. For holistic systems such as UniClean and Horizon, which do not explicitly output detected errors, we derive detection results from the repairs they perform: whenever a system changes a cell value, we count that cell as detected.

\subsubsection{Cleaning Effectiveness w.r.t.\ Error Rate.} Figure~\ref{fig:effectiveness} illustrates the effectiveness of all baselines in both error detection and correction on samples from \dsname{}, \deslice{}, and \ruslice{}. The \emph{x}-axis represents the cell-level error rate, while the \emph{y}-axis shows the F1-score.
As expected, all systems perform better on \deslice{} and \ruslice: within the same country, patterns and rules for addresses are more homogeneous, which makes the data inherently easier to clean.

\paragraph{Error Detection.} Among detection baselines, Raha shows the strongest performance across all slices. The LLM-based approach by Glock et al.\ is the second strongest detector. On \deslice{}, it reaches an F1-score of 0.77-0.82 and remains close to Raha across all error rates. DataWig, HoloClean, UniClean, and Horizon start from substantially lower F1-scores at small error rates, but their performance improves as the amount of errors increases. For example, on \deslice{} DataWig's F1-score rises from 0.13 at a 5\% error rate to 0.68 at 48\%. This trend can be explained by class imbalance. At low error rates, erroneous cells are rare. Thus, even high-recall detectors  may suffer from low precision, since many flagged cells are clean. As the error rate increases, the share of erroneous cells grows, precision improves, and the F1-score rises accordingly. This effect is less visible for Raha, which maintains competitive precision even at low error rates due to its semi-supervised strategy.

\paragraph{Error Correction.} To evaluate correction quality independently from detection, we provide the systems with the list of actual erroneous cells.
In contrast to detection, which benefits from class imbalance at higher error rates, correction performance degrades as the error rate increases.
Most correction systems rely on the assumption that the dataset contains enough clean and redundant evidence to infer the correct value. As the error rate increases, this evidence becomes less reliable, hence correction gets harder.

Baran achieves the strongest average performance across all slices. On \deslice{}, Baran and Glock's\ perform similarly, with average F1-scores of 69\% and 68\%, respectively. UniClean ranks third with 53\%.
A similar trend is registered on \ruslice{}, with Baran (45\%) outperforming UniClean (39\%) and Glock's\ (37\%).
The lower scores registered on \ruslice{} compared to \deslice{} are mainly due to the fact that its columns generally present higher distinct value ratios, providing therefore less evidence to support correction.
Finally, Baran reaches an average F1-score of 46\% on \dsname{}, followed by Glock's\ with 42\% and DataWig with 18\%.

Baran's strong performance shows the benefits of using signals within the dataset when enough clean context is available. Its performance drops for higher error rates, because the same internal signals become noisier as more cells are corrupted.
Glock's\ is more robust against increases in error rates because each record is processed independently and the method relies on knowledge acquired during pretraining rather than on evidence present in the dataset. DataWig also remains relatively stable, because we used a fixed training budget across all error rates.

Rule-based approaches perform poorly in correction because their applicability is limited by the available dependencies and by the values present in the data. FDs do not cover all attributes. Thus, errors in attributes such as \textsf{house\_number}, \textsf{name}, or \textsf{postcode} cannot be repaired, as they are not captured by any rule. Even when a rule applies, the system can only infer a correction if enough clean and consistent evidence exists elsewhere in the dataset. This assumption is often violated in our benchmark, which contains heterogeneous addresses from different countries, together with a high number of missing values, abbreviations, and noisy variants. These results show a limitation shared by many correction methods: they rely on the dataset itself to provide reliable correction evidence, hence their effectiveness decreases when the data is noisy, incomplete, or internally inconsistent.

\begin{table*}[t]
\centering
\caption{Error detection (recall) and correction (F1-score) performance across error types.}\label{tab:poi-err-type-res}
\vspace{-.3cm}

\scriptsize
\setlength{\tabcolsep}{3pt}
\renewcommand{\arraystretch}{1.0}
\begin{tabular}{c l l l cccccccccc}
\toprule
\textbf{Dataset} & \textbf{Task} & \textbf{Method}
& Missing & Parsing & Capitalization & Special Chars & Name Reformat & Language & Abbreviation & Token Subset & Typo & Word Subst. & Miscellaneous \\
\midrule

\multirow{12}{*}{\rotatebox[origin=c]{90}{\library{}}}
  & \multirow{5}{*}{\shortstack{Detection\\\textit{(Recall)}}}  & Raha         & 0.995 & 0.286 & 0.813 & 0.594 & 0.649 & 0.504 & 0.949 & 0.580 & 0.106 & 0.524 & 0.653 \\
  &   & Horizon      & 0.332 & \textbf{1.000} & \textbf{1.000} & \textbf{1.000} & \textbf{1.000} & \textbf{1.000} & 0.970 & \textbf{1.000} & \textbf{1.000} & \textbf{1.000} & \textbf{0.993} \\
  &   & DataWig      & 0.994 & 0.667 & 0.937 & 0.894 & 0.846 & 0.891 & \textbf{0.995} & 0.844 & 0.951 & 0.952 & 0.989 \\
  &   & HoloClean    & 0.959 & 0.000 & 0.137 & 0.033 & 0.142 & 0.102 & 0.390 & 0.223 & 0.244 & 0.000 & 0.395 \\
  &   & Glock's       & 0.995 & 0.000 & 0.003 & 0.000 & 0.000 & 0.013 & 0.061 & 0.006 & 0.081 & 0.143 & 0.175 \\
  &     & UniClean     & \textbf{1.000} & \textbf{1.000} & 0.551 & 0.683 & 0.544 & 0.570 & 0.501 & 0.579 & 0.512 & 0.571 & 0.626 \\

  \cmidrule{2-14}
  & \multirow{6}{*}{\shortstack{Correction\\\textit{(F1-Score)}}}  & Baran        & 0.072 & 0.000 & 0.043 & 0.065 & 0.019 & 0.172 & 0.712 & 0.041 & 0.089 & \textbf{0.250} & \textbf{0.427} \\
  &   & Horizon      & 0.010 & 0.000 & 0.000 & 0.000 & 0.000 & 0.000 & 0.014 & 0.000 & 0.000 & 0.000 & 0.000 \\
  &   & DataWig      & \textbf{0.399} & 0.000 & 0.022 & 0.000 & 0.016 & 0.018 & 0.154 & 0.008 & 0.008 & 0.000 & 0.016 \\
  &   & HoloClean    & 0.001 & 0.000 & 0.004 & 0.000 & 0.000 & 0.000 & 0.000 & 0.000 & 0.000 & 0.000 & 0.000 \\
  &   & Glock's       & 0.219 & 0.000 & \textbf{0.372} & \textbf{0.117} & \textbf{0.604} & \textbf{0.330} & \textbf{0.936} & \textbf{0.100} & \textbf{0.182} & 0.154 & 0.256 \\
  &   & UniClean    & 0.007 & 0.000 & 0.000 & 0.000 & 0.000 & 0.000 & 0.000 & 0.000 & 0.000 & 0.000 & 0.000 \\

\midrule

\multirow{12}{*}{\rotatebox[origin=c]{90}{\shoppingcenter{}}}
  & \multirow{6}{*}{\shortstack{Detection\\\textit{(Recall)}}}  & Raha         & \textbf{1.000} & 0.267 & 0.872 & 0.477 & 0.575 & 0.679 & 0.798 & 0.434 & 0.253 & 0.921 & 0.747 \\
  &   & Horizon      & 0.460 & \textbf{1.000} & \textbf{0.993} & \textbf{1.000} & \textbf{1.000} & \textbf{0.992} & 0.961 & \textbf{1.000} & \textbf{1.000} & \textbf{1.000} & 0.962 \\
  &   & DataWig      & 0.996 & 0.833 & 0.958 & 0.879 & 0.925 & 0.949 & \textbf{0.982} & 0.876 & 0.893 & 1.000 & \textbf{0.983} \\
  &   & HoloClean    & 0.996 & 0.000 & 0.268 & 0.000 & 0.226 & 0.101 & 0.152 & 0.028 & 0.200 & 0.000 & 0.506 \\
  &   & Glock's       & 0.999 & 0.000 & 0.001 & 0.023 & 0.000 & 0.009 & 0.061 & 0.013 & 0.027 & 0.000 & 0.169 \\
    &   & UniClean     & \textbf{1.000 }& \textbf{1.000} & 0.056 & 0.250 & 0.016 & 0.050 & 0.094 & 0.022 & 0.000 & 0.000 & 0.544 \\
  \cmidrule{2-14}
  & \multirow{6}{*}{\shortstack{Correction\\\textit{(F1-Score)}}}  & Baran        & \textbf{0.432} & 0.000 & \textbf{0.219} & 0.029 & \textbf{0.640} & 0.275 & 0.586 & 0.136 & 0.097 & 0.027 & \textbf{0.429} \\
  &   & Horizon      & 0.004 & 0.000 & 0.000 & 0.000 & 0.000 & 0.006 & 0.006 & 0.000 & 0.000 & 0.000 & 0.000 \\
  &   & DataWig      & 0.400 & 0.067 & 0.035 & 0.023 & 0.038 & 0.028 & 0.117 & 0.020 & 0.071 & 0.035 & 0.160 \\
  &   & HoloClean    & 0.001 & 0.000 & 0.023 & 0.000 & 0.000 & 0.000 & 0.000 & 0.000 & 0.000 & 0.000 & 0.000 \\
  &   & Glock's       & 0.210 & 0.000 & 0.129 & \textbf{0.300} & 0.225 & \textbf{0.446} & \textbf{0.762} & \textbf{0.148} & \textbf{0.105} & \textbf{0.563} & 0.344 \\
  &   & UniClean     & 0.008 & 0.000 & 0.000 & 0.000 & 0.000 & 0.005 & 0.000 & 0.000 & 0.000 & 0.000 & 0.016 \\
\bottomrule
\end{tabular}
\end{table*}

\subsubsection{Cleaning Effectiveness on POI-based Slices and Different Error Types} We evaluate the systems on two POI-based slices: \library{} (1,656 records, 13,248 cells) and \shoppingcenter{} (802 records, 6,416 cells).
These POIs were selected because they exhibit complementary error profiles and together cover the main error types identified in our benchmark, as shown in Table~\ref{tab:facility-error-summary}.

Table~\ref{tab:facility_results} reports the overall precision, recall, and F1-score for detection and correction on both slices. Table~\ref{tab:poi-err-type-res} further breaks down the results by error type. For detection, we report recall per error type, since the goal is to measure how many errors are successfully identified for each category. Type-wise precision is not meaningful in this context, because false positives correspond to clean cells and therefore cannot be assigned to a specific error category. For correction, we report F1-score, as a predicted repair is only counted as correct if it matches the ground-truth value.

For several error types, Raha, UniClean, DataWig, and Horizon achieve high detection recall, yet the corresponding correction F1-scores remain low. For example, on \library{} DataWig detects \textit{Abbreviation} errors with a recall of 0.995, but reaches only a 0.154 F1-score in correcting them. Similarly, Horizon obtains near-perfect recall for several categories, but almost never produces successful repairs. Rule violations or distributional signals are often enough to identify suspicious cells, but not to infer the correct value.

Table~\ref{tab:poi-err-type-res} shows that correction quality strongly depends on the error type. Baran performs best on \textit{Missing}, \textit{Name Reformat}, and \textit{Miscellaneous} errors on \shoppingcenter{}, where useful evidence can be derived from recurring patterns in the dataset. Glock's\ performs better on textual and semantic transformations, including \textit{Abbreviation}, \textit{Language}, \textit{Special Characters}, and \textit{Word Substitution}. This complements the previous results: learning-based and rule-based methods are effective when clean and redundant evidence is available, whereas the LLM-based method is more robust for transformations that require lexical, linguistic, or world knowledge.

\subsection{Efficiency and Scalability Experiments}

To evaluate efficiency and scalability of the different data cleaning systems, we measure runtime and peak memory usage. Table~\ref{tab:runtime_memory} shows the corresponding results. We executed all systems three times. The number of records ranges from 10\emph{k} to the full dataset size. Executions exceeding a predefined 10-hour limit were terminated.

\subsubsection{Runtime}
Among all baselines, DataWig is the fastest. Its runtime can be due to the small training set, consisting of only 20 tuples.
Both Raha and Baran were terminated before reaching 500\emph{k} records: the former ran out of memory, while the latter exceeded the time limit.
Although Horizon is usually fast, its runtime is sensitive to the number and interaction structure of the input FDs. Having to deal with 24 FDs at a 49\% error rate, the run did not finish within 10 hours on the 500\emph{k}-record dataset.

UniClean's runtime is non-monotonic. The execution on 100\emph{k} records (3,118s) is faster than on 10\emph{k} (13,455s) and 500\emph{k} (7,613s). UniClean is configured with a set of rules and repairs violations by first partitioning inconsistent rows into blocks. Cost scales with block size, rule count, and iteration count, so runtime is driven by error distribution rather than the number of rows.
Each sample presents a different error profile, determining the non-monotonic behavior mentioned above.
Finally, Glock's offloads computation to an external LLM API, avoiding local hardware limits, but one API call per record makes runtime and cost scale linearly, already requiring 8 hours to process 1M records and therefore exceeding the time limit on the full dataset with 4.3M records.
Clearly, also monetary costs become significant on large datasets.
For instance, just a single run on the full dataset would cost approximately \$650.
Note that Gemini 2.5 Flash-Lite is a rather fast model with a quick-response API. We had significantly slower runtime in preliminary experiments using GPT-5-mini.

\begin{table}[ht]
\scriptsize
\centering
\caption{Runtime (seconds) and peak memory usage (MB) of data cleaning systems across samples of increasing size.\\TO: timeout (runtime $>$ 10 hours); OOM: out of memory.}
\vspace{-.3cm}
\label{tab:runtime_memory}
\begin{tabular}{ll rrrrr}
\toprule
& & \multicolumn{5}{c}{\textbf{Number of Records}} \\
\cmidrule(lr){3-7}
\textbf{Task} & \textbf{System} & \textbf{10$k$} & \textbf{100$k$} & \textbf{500$k$} & \textbf{1M} & \textbf{4.3M} \\
\midrule
\multicolumn{7}{l}{\textit{Runtime (seconds)}} \\
\midrule
\multirow{3}{*}{Error Detection}
  & Raha      & 122.7  & 12865.9 & OOM & OOM & OOM \\
  & HoloClean & 8.5  & 45.6                        & 219.8 & 445.9 & 1,965.0 \\
  & Glock's   & 101.8  & 1092.9  & 10,379.46 & 12,927.58 & TO \\
\midrule
\multirow{3}{*}{Error Correction}
  & Baran     & 1,057.2  & 5,230.4 & TO & TO & TO \\

  & HoloClean & 404.5  & 4,621.1 & TO & TO & TO \\
  & Glock's   & 103.4  & 1,747.39     & 8,945.61 & 16,567.23 & TO \\
\midrule
\multirow{3}{*}{\shortstack[l]{Error Detection +\\Error Correction}}
 & DataWig   & 29.1  & 241.2  & 1,186.4 & 2,376.5 & 10,230.1 \\
 & Horizon   & 6,445.0  &  TO & TO & TO & TO \\
 & UniClean  &
13,455  & 3,118           & 7,613 & 11,357 & 13,490 \\
\midrule
\midrule

\multicolumn{7}{l}{\textit{Peak Memory (MB)}} \\
\midrule
\multirow{3}{*}{Error Detection}
  & Raha      & 1795  & 86,734 & OOM & OOM & OOM \\
  & HoloClean & 306  & 711                        & 2,459 & 4,663 & 18,979 \\
   & Glock's & 257.69 & 1,030.13 & 4,562.0 & 8,954.88 & TO \\
\midrule
\multirow{3}{*}{Error Correction}
  & Baran     & 685.84  & 3021.3  & TO & TO & TO \\
  & HoloClean & 2,867  & 33,843                        & TO & TO & TO \\
  & Glock's & 257.69 & 1,030.13 & 4,562.0 & 8,954.88 & TO \\
\midrule
\multirow{3}{*}{\shortstack[l]{Error Detection +\\Error Correction}}
& DataWig   & 1,138  & 2,002  & 3,723 & 5,761 & 14,576 \\
 & Horizon   & 1,700  & 2,088                        & TO & TO & TO \\
 & UniClean  & 8,229  & 	127,662     & 131,006 & 324,948 &
315,567 \\
\bottomrule
\end{tabular}
\end{table}

\subsubsection{Memory Usage}

Among the non-LLM approaches, DataWig is the most memory-efficient: it requires at most around 14~GB even on the full dataset. In contrast, Raha is substantially more memory-intensive. Its memory consumption increases from about 1.8~GB on the smallest setting to more than 86~GB on the subset with 100$k$ records, after which it runs out of memory. This limits its applicability to larger datasets despite its strong detection effectiveness. HoloClean also shows increasing memory usage, reaching around 19~GB for error detection on the full dataset. However, in correction settings it times out before memory becomes the limiting factor.
UniClean shows the highest memory footprint. Its peak memory already exceeds 127~GB on 100$k$ records and more than 300~GB on larger samples. Horizon, in comparison, remains relatively lightweight on the smaller samples, but it times out on larger ones.
Finally, the LLM-based approach has low memory requirements because the main computation is delegated to the cloud model. Thus, the reported peak memory usage does not reflect the actual memory footprint of the model.

\subsection{Key Takeaways}
Overall, our experiments show that even for address data, which is highly prevalent and does not require specialized domain expertise, existing cleaning systems remain insufficient for resolving data quality issues, prompting the need for further research on scalable and reliable data cleaning.
Learning-based methods are strong detectors and can exploit regularities within the dataset when sufficient clean evidence is available. LLM-based approaches are more robust under heterogeneity and missingness, but come with scalability costs. Rule-based systems are effective only in settings with high redundancy, reliable dependencies, and broad rule coverage.
In the following, we point out some more detailed takeaways.

\paragraph{(1) Data consistency.} Systems generally perform better on the country-specific slices than on the full dataset, confirming that linguistic, structural, and formatting consistency facilitates both detection and correction. Thus, a clear development direction for general-purpose cleaning systems is the ability to identify such slices automatically for effective step-wise cleaning.

\paragraph{(2) Error rates.} Detection often improves as the error rate increases, because erroneous cells become less rare and class imbalance becomes less severe. Correction, however, typically becomes harder at higher error rates, since the clean and redundant evidence needed to infer the correct value becomes less reliable. Future research should therefore consider including augmentation strategies to balance out the error rates.

\paragraph{(3) Cleaning strategies} We observe a clear tendency where supervised approaches outperform the more rigid rule-based systems, even on the rather canonical data types that exist in postal datasets. Although rule-based approaches are generally faster, their effectiveness depends on the error rate, dataset homogeneity, and availability of rules. While both rule-based and learning-based approaches primarily exploit patterns and evidence contained in the available data, LLMs can draw on knowledge acquired during pretraining to detect and correct erroneous values. This would suggest that a promising direction is a staged combination of these strategies, by filtering with fast and coarse-granular rule-based techniques and passing on difficult cases to the more powerful supervised and LLM-based solutions.
However, we have to acknowledge that existing systems already combine rules, supervision, and agents~\cite{DBLP:journals/pvldb/MahdaviA20,DBLP:conf/icde/NiZMZWWY25,DBLP:conf/icde/WuYZMNXZY25}. Thus, the way forward is to further refine and adjust how different approaches work together so that a quick fall-back on expensive agentic-based solutions can be avoided.

\paragraph{(4) Scalability} Existing systems scale differently and are therefore not equally suitable for all settings. Beyond table size, efficiency is affected by the error rate, the number and complexity of rules, and the distribution of errors and values in the data. LLM-based solutions avoid some local hardware limitations, but introduce a different scalability bottleneck: their cost and runtime depend on record-level inference, model response time, and the available serving infrastructure. Consequently, they can quickly become impractical at larger scales. Future cleaning research would benefit from approaches that take the relevant metadata into consideration and recalibrate their detection and correction strategies.

\subsubsection*{Limitations} While the dataset covers a very important domain and is large, it does not scale in the number of columns. Although current systems already struggle with the current dataset dimensions, it would be beneficial for future benchmarks to also scale horizontally.
Further, the dataset is skewed towards US and Europe.
Finally, the approach for obtaining ground truth is only one among many. Quality describes fitness for use. How addresses should be represented and used can differ depending on the use case. Our approach made sure that the data is correct and consistent with regard to one standard that was cross-confirmed with different validators.

\section{Conclusions}
\label{sec:conclusion}
In this paper, we present the first large-scale dirty address dataset with ground truth. The dirty version of the dataset contains real-world addresses. The collected addresses are dirty and inconsistent. The dataset contains a high variety of errors, enabling reasoning on different types and rates of errors. Using a geo-location service and additional manual postprocessing, we obtained a corresponding ground truth. For ease of use, we also provide country-based and POI-based slices of the dataset. With this dataset, we aim to support research in data cleaning. In fact, we also highlight that existing prototypes show significant limitations in handling our dataset.
Beyond data cleaning, the dataset might also be interesting for use cases such as data transformation discovery or entity resolution.

\bibliographystyle{ACM-Reference-Format}
\bibliography{abbreviations,references}

\end{document}